\documentclass[12pt]{iopart}

\usepackage{graphicx}

\begin{document}

\review[Introduction to half-metallic Heusler alloys]{Introduction to
half-metallic Heusler alloys: Electronic Structure and Magnetic
Properties}

\author{I Galanakis\dag, Ph Mavropoulos\ddag\ and P H Dederichs\ddag}

\address{\dag\ Materials Science Department, School of Natural
  Sciences, University of Patras, Patras 265 04, Greece}
\address{\ddag\ Institut f\"ur Festk\"orperforschung, Forschungszentrum J\"ulich, D-52425
J\"ulich, Germany}

\ead{i.galanakis@fz-juelich.de,ph.mavropoulos@fz-juelich.de,p.h.dederichs@fz-juelich.de}

\begin{abstract}
  Intermetallic Heusler alloys are amongst the most attractive
  half-metallic systems due to the high Curie temperatures and the
  structural similarity to the binary semiconductors. In this review
  we present an overview of the basic electronic and magnetic
  properties of both Heusler families: the so-called half-Heusler
  alloys like NiMnSb and the the full-Heusler alloys like Co$_2$MnGe.
  \textit{Ab-initio} results suggest that both the electronic and
  magnetic properties in these compounds are intrinsically related to
  the appearance of the minority-spin gap.  The total spin magnetic
  moment $M_t$ scales linearly with the number of the valence
  electrons $Z_t$, such that $M_t=Z_t-24$ for the full-Heusler and
  $M_t=Z_t-18$ for the half-Heusler alloys, thus opening the way to
  engineer new half-metallic alloys with the desired magnetic
  properties.
\end{abstract}

\pacs{ 75.47.Np, 71.20.Be, 71.20.Lp}

\submitto{\JPD}

\maketitle


\section{Half-metallic Heusler Alloys \label{secios:1}}
\begin{figure}
\centering
\includegraphics[width=\linewidth]{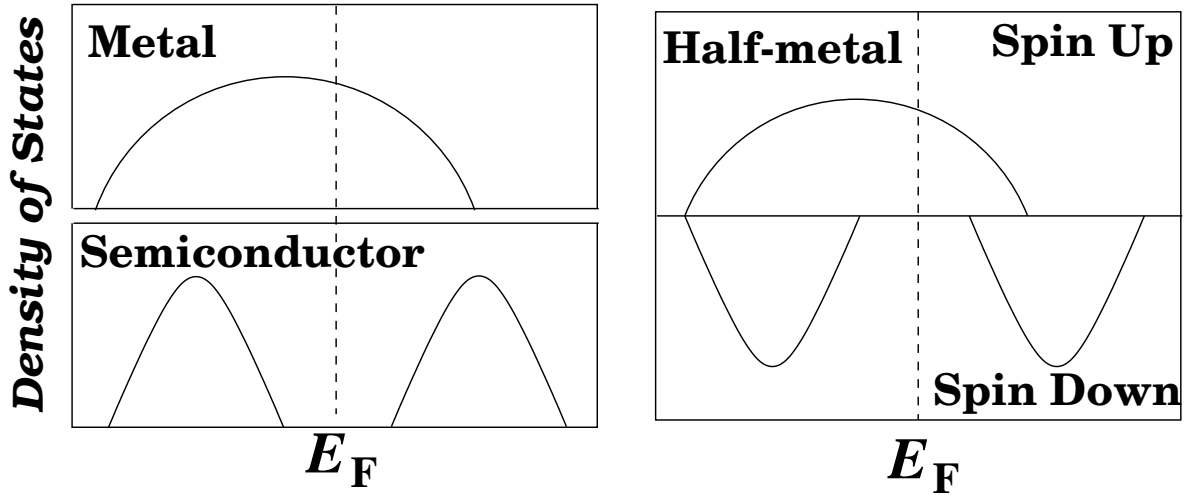}
\caption{Schematic representation of the density of states for a
half-metal with respect to normal metals and semiconductors.}
\label{figios1a}
\end{figure}

Half-metallic ferromagnets represent a relatively new class of
materials which have attracted recently a lot of interest due to their
possible applications in spin electronics (also known as
magnetoelectronics)~\cite{Zutic2004}. In these materials the two spin
bands show a completely different behaviour. While one of them
(usually the majority-spin band, henceforth also refered to as spin-up
band) shows a typical metallic behaviour with a nonzero density of
states at the Fermi level $E_F$, the minority (spin-down) band
exhibits a semiconducting behaviour with a gap at $E_F$. Therefore
such half-metals can be considered as hybrides between metals and
semiconductors. A schematic representation of the density of states of
a normal metal, a semiconductor and a half-metal is shown in
figure~\ref{figios1a}. Such half-metals exhibit, ideally, a 100\% spin
polarization at the Fermi level and therefore these compounds should
have a fully spin-polarized current and be ideal spin injectors into a
semiconductor, thus maximizing the efficiency of spintronic
devices~\cite{Wolf}.

\begin{figure}
\centering
\includegraphics[width=\linewidth]{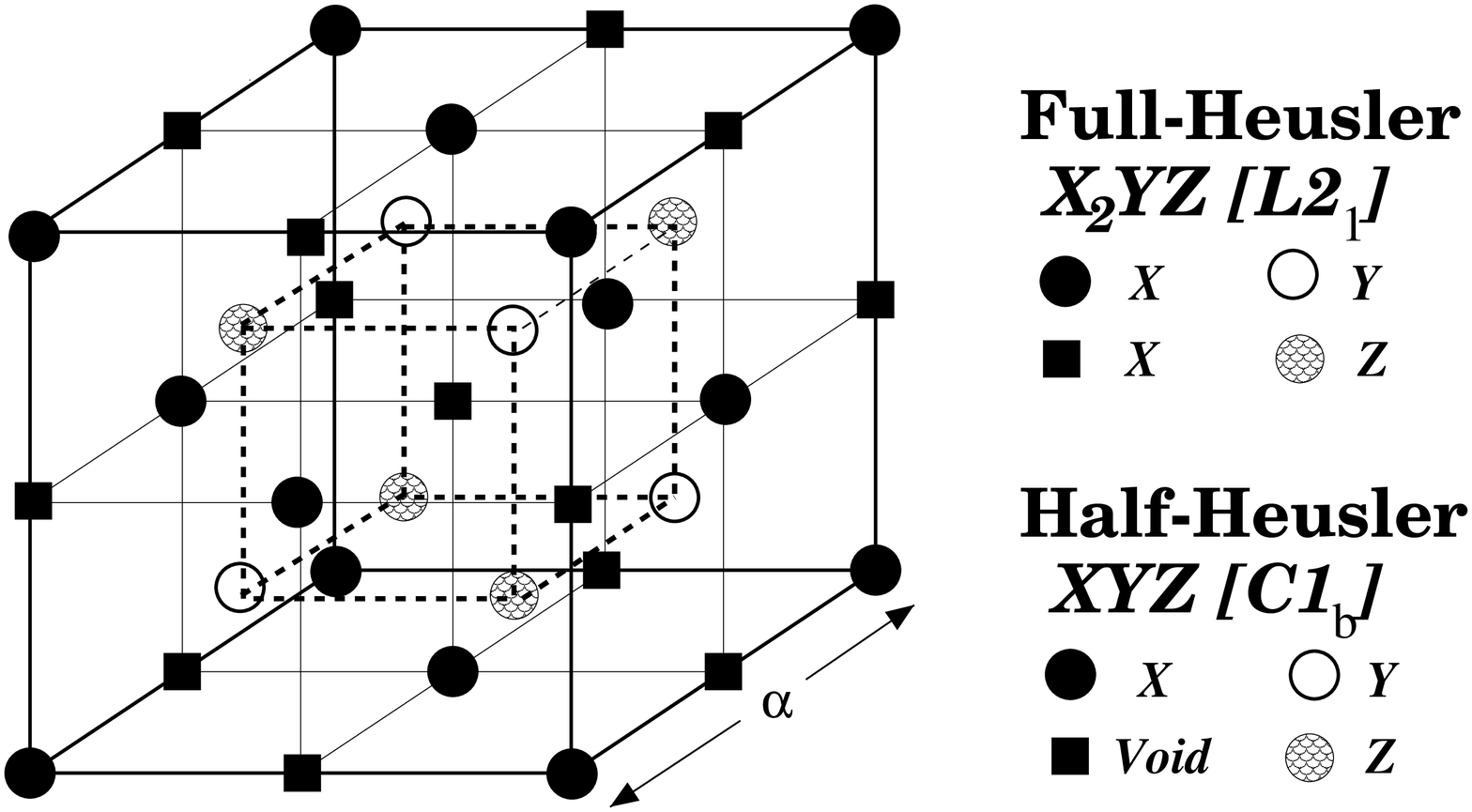}
\caption{$C1_b$ and $L2_1$ structures adapted by the half- and
full-Heusler alloys. The lattice is consisted of 4
interpenetratating fcc sublattices. The unit cell is that of a fcc
lattice with four atoms as basis, \textit{e.g.} CoMnSb: Co at
$(0\:0\:0)$, Mn at $({1\over4}\:{1\over4}\:{1\over4})$, a vacant
site at $({1\over2}\:{1\over2}\:{1\over2})$  and Sb at
$({3\over4}\:{3\over4}\:{3\over4})$  in Wyckoff coordinates. In
the case of the full Heusler alloys also the vacant site is
occupied by a Co atom. Note also that if all atoms were identical,
the lattice would be simply the bcc.} \label{figios1}
\end{figure}

Heusler alloys~\cite{heusler} have attracted great interest during the
 last century due to the possibility to study in the same family of
 alloys a series of interesting diverse magnetic phenomena like
 itinerant and localized magnetism, antiferromagnetism, helimagnetism,
 Pauli paramagnetism or heavy-fermionic behavior
 \cite{landolt,landolt2,Pierre97,Tobola}. Recently, also their
 application as shape-memory alloys has been intensively
 discussed~\cite{Jussi}. The first Heusler alloys studied were of the
 form X$_2$YZ and crystalize in the L2$_1$ structure which consists of
 four fcc sublattices, two of which are occupied by the same type of
 X-atoms (see figure~\ref{figios1}). Afterwards the XYZ Heusler alloys
 of C1$_b$ structure were discovered, where one sublattice remains
 unoccupied. The latter compounds are often called half- or
 semi-Heusler alloys, while the L2$_1$ compounds are referred to as
 full Heusler alloys.

In 1983 de~Groot and co-workers~\cite{groot} discovered by {\it
ab-initio} calculations that one of the half-Heusler alloys, NiMnSb,
is half-metallic, i.e., the minority band is semiconducting with a
band gap at $E_F$, as shown in figure~\ref{figios2}. Since then, many
other materials have also been found to be half-metallic. Besides a
number of half- and full-Heusler alloys, these are some oxides
(\textit{e.g.}, CrO$_2$ and Fe$_3$O$_4$) \cite{Soulen98}, manganites
(\textit{e.g.}, La$_{0.7}$Sr$_{0.3}$MnO$_3$) \cite{Soulen98}, double
perovskites (\textit{e.g.}, Sr$_2$FeReO$_6$) \cite{Kato}, pyrites
(\textit{e.g} CoS$_2$) \cite{Pyrites}, transition metal chalcogenides
(\textit{e.g.}, CrSe) and pnictides (\textit{e.g} CrAs) in the
zinc-blende or wurtzite structures
\cite{GalanakisZB,Xie,Akinaga,Zhao}, europium chalcogenides
(\textit{e.g.}, EuS) \cite{Temmerman}, and diluted magnetic
semiconductors (\textit{e.g} Mn impurities in Si or
GaAs)\cite{FreemanMnSi,Akai98}.  Although thin films of CrO$_2$ and
La$_{0.7}$Sr$_{0.3}$MnO$_3$ have been verified to present practically
100\% spin-polarization at the Fermi level at low temperatures
\cite{Soulen98,Park98}, the Heusler alloys remain attractive for
technological applications like spin-injection devices \cite{Datta},
spin-filters \cite{Kilian00}, tunnel junctions \cite{Tanaka99}, or GMR
devices \cite{Caballero98} due to their relatively high Curie
temperatures compared to other half-metallic compounds \cite{landolt}.

The half-metallic character of NiMnSb in single crystals seems to have
been well-established experimentally. Infrared absorption
\cite{Kirillova95} and spin-polarized positron-annihilation
\cite{Hanssen90} gave a spin-polarization of $\sim$100\% at the Fermi
level.  However, in many other experiments half-metallicity has not
been found. In most cases the reason for this is that properties have
been measured which are surface and interface sensitive. It has been
shown \cite{picozziMult,GalanakisInter2,ShiraiInter} that near
surfaces and interfaces the half-metallic property is in most cases
lost.  Therefore most of these experiments do not allow for
conclusions about half-metallicity in the bulk to be drawn.

In this introductory review we discuss the basic electronic and
magnetic properties of the half-metallic Heusler alloys. Analyzing
\textit{ab-initio} results and using group theory and simple models we
explain the origin of the gap in both the half- and full-Heusler
alloys as arising from $d$-$d$ hybridisation, which is fundamental for
understanding their electronic and magnetic properties. For both
families of compounds the total spin magnetic moment scales with the
number of valence electrons and can be described by a Slater-Pauling
rule, thus opening the way to engineer new half-metallic Heusler
alloys with desired magnetic properties. This behaviour is governed by
the fact that in these alloys the number of minority valence electrons
per unit cell is an integer given by the number of occupied minority
valnce bands, being equal to 9 for the half-Heusler and 12 for the
full-Heusler compounds. We discuss the role of spin-orbit coupling,
which in principle destroys the minority band gap, however in practice
leads only to a small reduction of spin polarization. We will not
discuss the effects of defects, interfaces or surfaces, for which we
refer to a recent book~\cite{Springer} and to many papers in this
issue.

In the following sections we discuss the electronic and magnetic
properties of half-metallic half-Heusler compounds XYZ
(Sect.~\ref{secios:2}) and of full-Heusler compounds X$_2$YZ
(Sect.~\ref{secios:3}), in particular the origin of the minority gap
and the resulting Slater-Pauling rules. Sect.~\ref{sect:4} shortly
addresses the effect of the lattice parameter, while in
Sect.~\ref{sect:5} the effect of spin-orbit coupling is investigated,
reducing the spin polarization at $E_F$. We conclude with a summary in
Sect.~\ref{secios:6}.

\section{Electronic and Magnetism Properties of Half-Heusler Alloys}
\label{secios:2}

\subsection{Band Structure of Half-Heusler Alloys}
\label{secios:2-1}

\begin{figure}
\centering
\includegraphics[width=\linewidth]{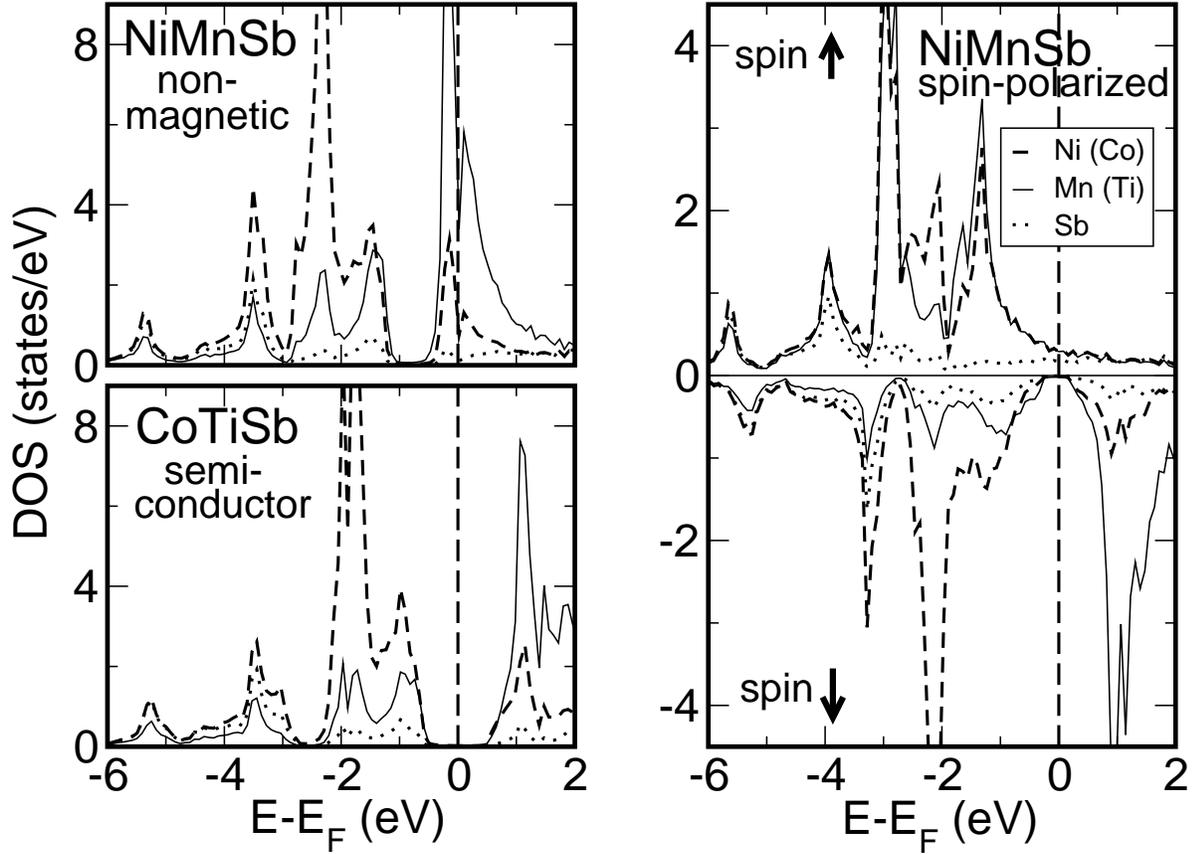}
\caption{Atom-resolved density of states (DOS) of NiMnSb for a
paramagnetic (upper left) and ferromagnetic (right) calculation.
In the left bottom panel the DOS for the semiconductor CoTiSb. The
zero energy value corresponds to the Fermi level $E_\mathrm{F}$.}
\label{figios2}
\end{figure}

In the following we present results for some typical half-Heusler
alloys of C1$_b$ structure (see figure \ref{figios1}).  To perform the
calculations, we used density-functional theory in the local density
approximation (LDA) in connection with the full-potential screened
Korringa-Kohn-Rostoker (FSKKR) method \cite{Zeller95,Papanikolaou02}.
The prototype example is NiMnSb, the half-metal discovered in 1983 by
de Groot \cite{groot}. Figure~\ref{figios2} shows the density of
states (DOS) of NiMnSb in a non-spin-polarized calculation (left upper
panel) and in a calculation correctly including the spin-polarization
(right panel). Given are the local contributions to the density of
states (LDOS) on the Ni site (dashed), the Mn site (full line) and the
Sb site (dotted). In the non-magnetic case the DOS of NiMnSb has
contributions from 4 different bands: Each Sb atom with the atomic
configuration 5$s^2$5$p^3$ introduces a deep lying $s$ band, which is
located at about $-12$eV and is not shown in the figure, and three Sb
$p$-bands in the regions between $-5.5$ and $-3$~eV. These bands are
separated by a deep minimum in the DOS from 5 Ni $d$ bands between $-3$
and $-1$ eV, which themselves are separated by a sizeable band gap from
the upper 5 $d$-bands of Mn. Since all atomic orbitals, {\it i.e.},
the Ni $d$, the Mn $d$ and the Sb $sp$ orbitals hybridise with each
other, all bands are hybrids between these states, being either of
bonding or antibonding type. Thus the Ni $d$-bands contain a bonding
Mn $d$ admixture, while the higher Mn $d$-bands are antibonding
hybrids with small Ni $d$-admixtures. Similarly, the Sb $p$-bands
exhibit strong Ni $d$- and somewhat smaller Mn $d$-contributions.

\begin{figure}
\centering
\includegraphics*[scale=0.6]{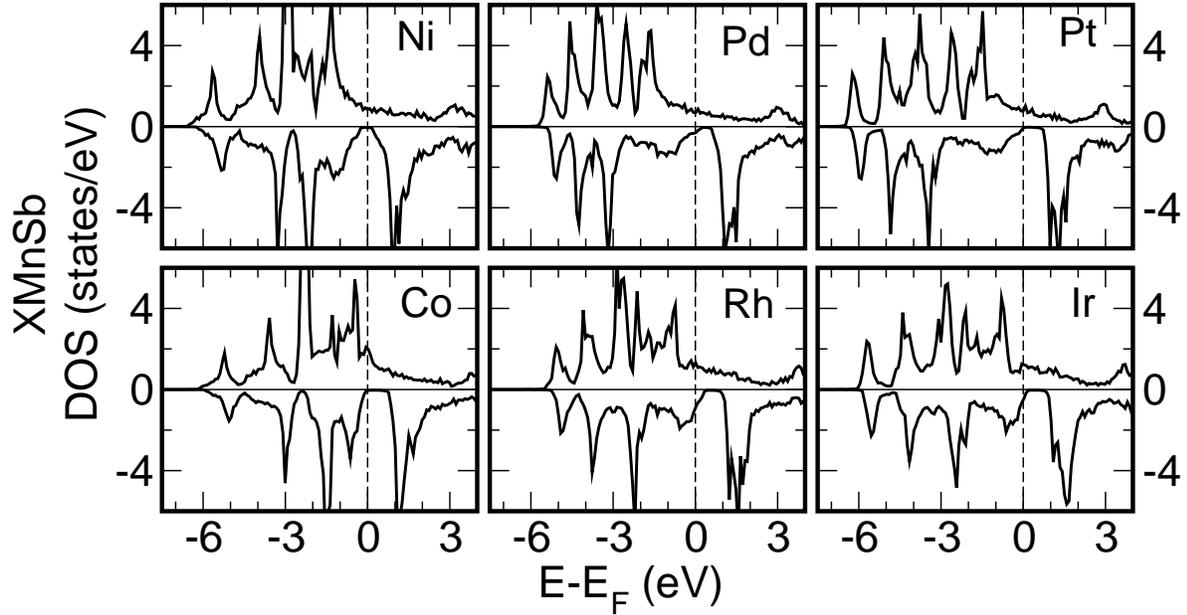}
\caption{DOS of XMnSb compounds for X= Ni, Pd, Pt and Co, Rh, Pd} \label{figios3}
\end{figure}

This configuration for NiMnSb is energetically not stable, since (i)
the Fermi energy lies in the middle of an antibonding band and (ii)
since the Mn atom can gain considerable exchange energy by forming a
magnetic moment. Therefore the spin-polarized results
(figure~\ref{figios2} right) show a considerably different
picture. In the majority (spin $\uparrow$) band the Mn $d$ states are
shifted to lower energies and form a common $d$ band with the Ni $d$
states, while in the minority band (spin $\downarrow$) the Mn states
are shifted to higher energies and are unoccupied, so that a band gap
at $E_F$ is formed separating the occupied $d$ bonding from the
unoccupied $d$-type antibonding states. Thus NiMnSb is a half-metal,
with a band gap at $E_F$ in the minority band and a metallic $sp$-like
DOS at $E_F$ in the majority band. The total magnetic moment per unit
cell, located mostly at the Mn atom, can be easily estimated to be
exactly 4 $\mu_B$. Note that NiMnSb has 22 valence electrons per unit
cell, 10 from Ni, 7 from Mn and 5 from Sb. Since, due to the gap at
$E_F$, in the minority band exactly 9 bands are fully occupied (1
Sb-like $s$ band, 3 Sb-like p bands and 5 Ni-like $d$ bands) and
accommodate 9 electrons per unit cell, the majority band contains $22 -
9 = 13$ electrons, resulting in the moment of 4 $\mu_B$ per unit cell.

The above non-spinpolarized calculation for NiMnSb (figure
\ref{figios2} left upper panel) suggests, that if we could shift
the Fermi energy as in a rigid band model, a particular stable
compound would be obtained if the Fermi level falls for both spin
directions into the band gap. Then for both spin directions 9
bands would be occupied, resulting in a semiconductor with 18
valence electrons. Such semiconducting Heusler alloys indeed
exist. As a typical example, figure \ref{figios2} shows also the
DOS of CoTiSb, which has a gap of 0.82 eV \cite{Tobola}. The gap is
indirect corresponding to transitions from the valence band
maximum at $\Gamma$ to the conduction band minimum at $X$. Other
such semiconductors are CoZrSb (0.83~eV), FeVSb (0.36~eV) and
NiTiSn (0.14~eV), where the values in the bracket denote the size
of the gap~\cite{Tobola}.

The band structure of the minority states of the half-Heusler alloys
is most important for the understanding of the magnetic properties,
and this band stucture is basically universally valid for all
half-Heusler alloys including the above semiconductors.  This can be
seen, for example, in the work of Nanda and Dasgupta who studied the
band structure of the half-Heusler semiconductor
FeVSb~\cite{Nanda03}. There, a ``fat band representation'' was used,
with the plotted widths of the bands indicating the weight of the
local orbitals to the band eigenstates. It was shown that around the
band gap the eigenstates are dominated by the Fe and V $d$ states
which strongly hybridise with each other, with the Fe weights being
stronger in the valence $d$ band and the V states in the conduction
band. The analogous feature can also be seen in the local density of
states in figure~\ref{figios2}. Below the five Fe $d$ bands are the
three Sb $p$ bands which have also small contributions of Fe and V
states. Analogously the Fe and V bands have small contributions of Sb
$p$ states (and even smaller contributions from the Sb $s$
states). The V $t_{2g}$ and $e_g$ states exhibit a small crystal field
splitting of 0.7~eV, with the $e_g$ level being lower due to their
weaker hybridisation with the Fe $e_g$ states in tetrahedral symmetry,
while for the Fe states this splitting is negligible. Most important
is, that the gap arises from the hybridisation of the Fe and the V
states and that the Sb $p$ states are very deep and do not affect the
gap.

\begin{table}
\centering \caption{Calculated spin magnetic moments in $\mu_B$
for the XMnSb compounds. (The experimental lattice constants
\protect{\cite{landolt}} have been used.) The deviations from integral
total moment for the half-metallic compounds are due to numerical
angular momentum cutoff (see text). \label{tableios:1}}
\begin{tabular}{r|r|r|r|r|r|c}
\hline\noalign{\smallskip}
 $m^{spin}(\mu_B)$ & X & Mn & Sb & Void
& Total & Half-metallic\\ \noalign{\smallskip}\hline\noalign{\smallskip}
 NiMnSb & 0.264 & 3.705 & -0.060 & 0.052 & 3.960 & yes\\
PdMnSb & 0.080 & 4.010 & -0.110 & 0.037 &4.017 & no\\ PtMnSb & 0.092 &
3.889 & -0.081 &
0.039 &3.938& no\\ CoMnSb & -0.132 & 3.176 & -0.098 & 0.011& 2.956 & yes\\
RhMnSb & -0.134 & 3.565 & -0.144 & $<$0.001 & 3.287 & no\\ IrMnSb &
-0.192 & 3.332 & -0.114 & -0.003 & 3.022 & no\\ FeMnSb & -0.702 &
2.715 & -0.053 & 0.019  & 1.979 & yes\\ \noalign{\smallskip}\hline
\end{tabular}
\end{table}

Table~\ref{tableios:1} summarises the calculated magnetic moments for
a series of XMnSb-type half-Heusler alloys. The calculations of the
density of states show that NiMnSb, CoMnSb and FeMnSb are half-metals,
{\it i.e.}, the Fermi level is in the minority gap, while for PtMnSb,
PdMnSb and IrMnSb $E_F$ enters slightly and for RhMnSb more strongly
into the valence band, reducing the spin polarization.

The total magnetic moment in $\mu_B$ is just the difference between
the number of occupied spin-up states and occupied spin-down states.
The number of spin-down bands below the gap is in all cases
$N_{\downarrow}=9$. Ignoring for a moment the small penetration of
$E_F$ into the valence band for some of these compounds, we directly
deduce the number $N_{\uparrow} = Z_t - N_{\downarrow} = Z_t - 9$ of
occupied spin-up states and the moment,
$M=(N_{\uparrow}-N_{\downarrow})\ \mu_B = (Z_t - 2\times
N_{\downarrow})\ \mu_B = (Z_t - 18)\ \mu_B$ from the total number of
valence electrons, $Z_t$. We then find $N_{\uparrow}=13$ and $M=4\
\mu_B$ for NiMnSb and for the isovalent compounds with Pd and Pt,
$N_{\uparrow}=12$ and $M=3\ \mu_B$ for CoMnSb, RhMnSb and IrMnSb, and
$N_{\uparrow}=11$ and $M=2\ \mu_B$ for FeMnSb, provided that the Fermi
level stays within the gap. Penetration of $E_F$ into the valence band
for PtMnSb, PdMnSb, IrMnSb, and RhMnSb causes a deviation of $M$ from
the integral values.

The {\it ab-initio} calculated local moment per unit cell as given in
table \ref{tableios:1} is close to 4 $\mu_B$ in the case of NiMnSb,
PdMnSb and PtMnSb, which is in agreement with the half-metallic
character (or nearly half-metallic character in the case of PdMnSb)
observed in the calculations. Note that due to the angular momentum
cutoff the KKR method can only give the correct integer number 4, if
Lloyd's formula is used in the evaluation of the integrated density of
states \cite{Rudi}, which is not the case in the present
calculations. We also find that the local moment of Mn is not far away
from the total number of 4 $\mu_B$ although there are significant
(positive) contributions from the X-atoms and a negative contribution
from the Sb atom. The antiferromagnetic coupling between the Sb and Mn
moments is due to the different behavior of the $p$ bands created by
the Sb atoms, discussed in the next section. The minority $p$ bands are
more located within the Wigner-Seitz cell of Sb. On the other hand the
majority $p$ bands are more expanded in space and contain a larger Mn
$d$-admixture and thus the total Sb spin moment has a negative sign.

We also find that for the half-metallic CoMnSb and IrMnSb compounds
the total moment is about 3 $\mu_B$. Here, the local moment of Mn is
higher than the total moment by at most 0.5 $\mu_B$. The reduction of
the total moment to 3 $\mu_B$ is therefore accompanied by negative Co
and Ir spin moments, \textit{i.e.}, these atoms couple
antiferromagnetically to the Mn moments.  The hybridization between Co
and Mn is considerably larger than between Ni and Mn, and is a
consequence of the smaller electronegativity difference and the larger
extent of the Co orbitals. Therefore the minority valence band of
CoMnSb has a larger Mn admixture than the one of NiMnSb whereas the
minority conduction band of CoMnSb has a larger Co admixture than the
Ni admixture in the NiMnSb conduction band, while the populations of
the majority bands are barely changed. As a consequence, the Mn moment
is reduced by the increasing hybridization, while the Co moment
becomes negative, resulting finally in a reduction of the total moment
from 4 to 3 $\mu_B$. Table \ref{tableios:1} also shows that
substitution of Co with Fe leads again to a half-metallic alloy with a
total spin moment of 2 $\mu_B$ as has been already shown by de Groot
\textit{et al.} in reference~\cite{Groot86}.

\subsection{Origin of the Gap\label{secios:2-3}}

\begin{figure}
\centering
\includegraphics[height=5cm]{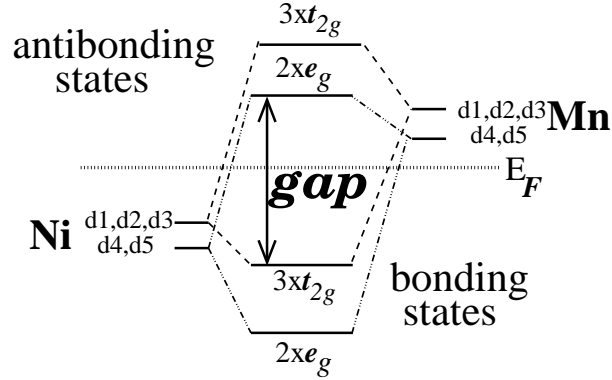}
\caption{Schematic illustration of the origin of the gap in the
minority band in half-Heusler alloys and semiconductors: The energy
levels $E_b$ of the energetically lower lying bonding hybrids are
separated from the levels $E_{ab}$ of the antibonding hybrids by a
gap, such that only the bonding states are occupied. Due to legibility
reasons, we use d1, d2 and d3 to denote the $d_{xy}$, $d_{yx}$ and
$d_{zx}$ $t_{2g}$-orbitals, respectively, and d4, d5 for the $d_{z^2}$,
$d_{x^2-y^2}$ $e_g$-orbitals.} \label{figios4}
\end{figure}

The local DOS shown in figure~\ref{figios2} for the ferromagnet NiMnSb
and for the semiconductor CoTiSb show that the DOS close to the gap is
dominated by $d$-states: in the valence band by bonding hybrids with
large Ni or Co admixture and in the conduction band by the antibonding
hybrids with large Mn or Ti admixture. Thus the gap originates from
the strong hybridization between the $d$ states of the two transition
metal atoms. This is sketched schematically in figure
\ref{figios4}. In this respect the origin of the gap is similar to the
gap in compound semiconductors like GaAs which is enforced by the
hybridization of the lower lying As $sp$-states with the energetically
higher Ga $sp$-states. Note that in the C1$_b$-structure the Ni and Mn
sublattices form a zinc-blende structure, which is important for the
formation of the gap. The difference with respect to GaAs is then that
5 $d$-orbitals, i.e. 3 $t_{2g}$ and 2 $e_g$ orbitals, are involved in
the hybridization, instead of 4 $sp^3$-hybrids in the compound
semiconductors.

\begin{figure}
\centering
\includegraphics[scale=0.6]{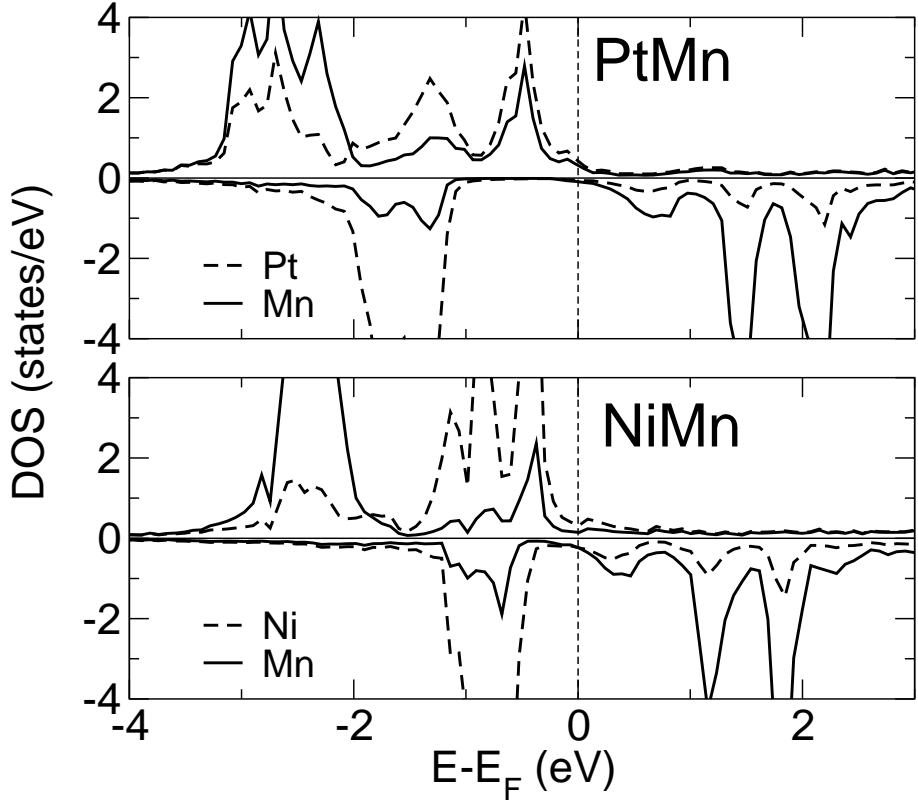}
\caption{Atom-resolved DOS for the hypothetical PtMn and NiMn
crystallizing in the zinc-blende structure.} \label{figios5}
\end{figure}

Giving these arguments it is tempting to claim, that also a
hypothetical zinc-blende compound like NiMn or PtMn should show a
half-metallic character with a gap at $E_F$ in the minority
band. Figure \ref{figios5} shows the results of a self-consistent
calculation for such zinc-blende NiMn and PtMn, with the same lattice
constant as NiMnSb. Indeed a gap is formed in the minority band. In
the hypothetical NiMn the Fermi energy is slightly above the gap,
however the isoelectronic PtMn compound shows nearly half-metallicity.
In this case the occupied minority states consist of six bands, a
low-lying $s$-band and five bonding $d$-bands, both of mostly Pt
character.  Since the total number of valence electrons is 17, the
majority bands contain 11 electrons, so that the total moment per unit
cell is $11\:-\: 6\:=\: 5 \mu_B$, which is indeed obtained in the
calculations. This is the largest possible moment for this compound,
since in the minority band all 5 Mn $d$-states are empty while all
majority $d$-states are occupied. The same limit of $5 \mu_B$ is also
the maximal possible moment of the half-metallic $C1_b$ Heusler
alloys.

The gap in the half-metallic $C1_b$ compounds is normally indirect,
with the maximum of the valence band at the $\Gamma$ point and the
minimum of the conduction band at the $X$-point. For NiMnSb we obtain
a band gap of about 0.5 eV, which is in good agreement with the
experiments of Kirillova and collaborators \cite{Kirillova95}, who,
analyzing their infrared spectra, estimated a gap width of $\sim 0.4$
eV. As seen already from figure \ref{figios3} the gap of CoMnSb is
considerable larger ($\sim 1$ eV) and the Fermi level is located at
the edge of the minority valence band.

As it is well-known, the local density approximation (LDA) and the
generalized gradient approximation (GGA) strongly underestimate the
values of the gaps in semiconductors, typically by a factor of
two. However, very good values for these gaps are obtained in the
so-called GW approximation of Hedin and Lundqvist \cite{Hedin}, which
describes the screening in semiconductors very well. On the other hand
the minority gap in the half-metallic systems might be better
described by the LDA and GGA since in these systems the screening is
metallic.

\subsection{Role of $sp$-Elements}
\label{secios:2-4}

While the $sp$-elements are not responsible for the existence of the
minority gap, they are nevertheless very important for the physical
properties of the Heusler alloys and the structural stability of the
$C1_b$ structure, as we discuss in the following. There are three
important features:

(i) While an Sb atom has 5 valence electrons (5$s^2$, 5$p^3$), in the
NiMnSb compound each Sb atom introduces a deep lying $s$-band, at
about $-12$ eV, and three $p$-bands below the center of the
$d$-bands. These bands accommodate a total of 8 electrons per unit
cell, so that formally Sb acts as a triple charged Sb$^{3-}$ ion.
Analogously, a Te-atom behaves in these compounds as a Te$^{2-}$
ion and a Sn-atom as a Sn$^{4-}$ ion. This does not mean, that
locally such a large charge transfer exists. In fact, the $s$- and
$p$-states strongly hybridize with the TM $d$-states and the
charge in these bands is delocalized and locally Sb even loses
about one electron, if one counts the charge in the Wigner-Seitz
cells. What counts here is that the $s$- and $p$-bands accommodate 8
electrons per unit cell, thus effectively reducing the $d$-charge
of the TM atoms.

This is nicely illustrated by the existence of the semiconducting compounds
CoTiSb and NiTiSn. Compared to CoTiSb, in NiTiSn the missing $p$-charge of
the Sn atom is replaced by an increased $d$ charge of the Ni atom, so that
in both cases all 9 valence bands are occupied.

(ii) The $sp$-atom is very important for the structural stability of
the Heusler alloys. For instance, it is difficult to imagine that
the calculated half-metallic NiMn and PtMn alloys with zinc-blende
structure, the LDOS of which are shown in figure \ref{figios5},
actually  exist, since metallic alloys prefer highly coordinated
structures like fcc, bcc, hcp etc. Therefore the $sp$-elements are
decisive for the stability of the $C1_b$ compounds. A careful
discussion of the bonding in these compounds has been recently
published by Nanda and Dasgupta \cite{nanda-dasgupta} using the
crystal orbital Hamiltonian population (COHP) method. For the
semiconductor FeVSb they find that, while the largest contribution
to the bonding arises from the V-$d$ -- Fe-$d$ hybridization,
contributions of similar size arise also from the Fe-$d$ -- Sb-$p$
and the V-$d$ -- Sb-$p$ hybridization. Similar results are also
valid for the semiconductors like CoTiSb and NiTiSn and in
particular for the half-metal NiMnSb. Since the majority $d$-band
is completely filled, the major part of the bonding arises from
the minority band, so that similar arguments as for the
semiconductors apply.

(iii) Another property of the $sp$-elements is worthwhile mentioning:
substituting the Sb atom in NiMnSb by Sn, In or Te destroys the
half-metallicity \cite{Galanakis2002a}. This is in contrast to the
substitution of Ni by Co or Fe, which is documented in table
\ref{tableios:1}. The total moment of 4 $\mu_B$ for NiMnSb is reduced
to 3 $\mu_B$ in CoMnSb and 2 $\mu_B$ in FeMnSb, thus preserving
half-metallicity. In NiMnSn the total moment is reduced only to 3.3
$\mu_B$ (instead of 3) and in NiMnTe the total moment increases only
to 4.7 $\mu_B$ (instead of 5). Thus by changing only the $sp$-element
it is rather difficult to preserve the half-metallicity, since the
density of states changes more like in a rigid band model
\cite{Galanakis2002a}.

\subsection{Slater-Pauling Behavior}
\label{secios:2-5}

As discussed above the total moment of the half-metallic $C1_b$
Heusler alloys follows the simple rule: $M_t = Z_t - 18$, where $Z_t$
is the total number of valence electrons per unit cell. In short, the
total number of electrons $Z_t$ is given by the sum of the number of
spin-up and spin-down electrons, while the total moment $M_t$ per unit
cell is given by the difference
\begin{equation}
Z_t = N_\uparrow + N_\downarrow \quad , \quad
M_t = N_\uparrow - N_\downarrow \quad \to \quad
M_t = Z_t - 2N_\downarrow
\end{equation}
Since 9 minority bands are fully occupied, we obtain the simple ''rule of 18''
for half-metallicity in $C1_b$ Heusler alloys
\begin{equation}
M_t = Z_t - 18
\end{equation}
the importance of which has been recently pointed out by Jung et
al. \cite{jung} and Galanakis et al. \cite{Galanakis2002a}. It is a
direct analogue to the well-known Slater-Pauling behavior of the
binary transition metal alloys \cite{Kubler84}.  The difference with
respect to these alloys is, that in the half-Heusler alloys the
minority population is fixed to 9, so that the screening is achieved
by filling the majority band, while in the transition metal alloys the
majority band is filled with 5 $d$-states and charge neutrality is
achieved by filling the minority states. Therefore in the TM alloys
the total moment is given by $M_t = 10 - Z_t$. Similar rules with
integer total moments are also valid for other half-metals, e.g. for
the full-Heusler alloys like Co$_2$MnGe with $L2_1$ structure. For
these alloys we will in section 3 derive the ``rule of 24'': $M_t =
Z_t - 24$, with arises from the fact that the minority band contains
12 electrons. For the half-metallic zinc-blende compounds like CrAs
the rule is: $M_t = Z_t - 8$, since the minority As-like valence bands
accommodate 4 electrons \cite{GalanakisZB}. In all cases the moments
are integer.

\begin{figure}
\centering
\includegraphics[height=7cm]{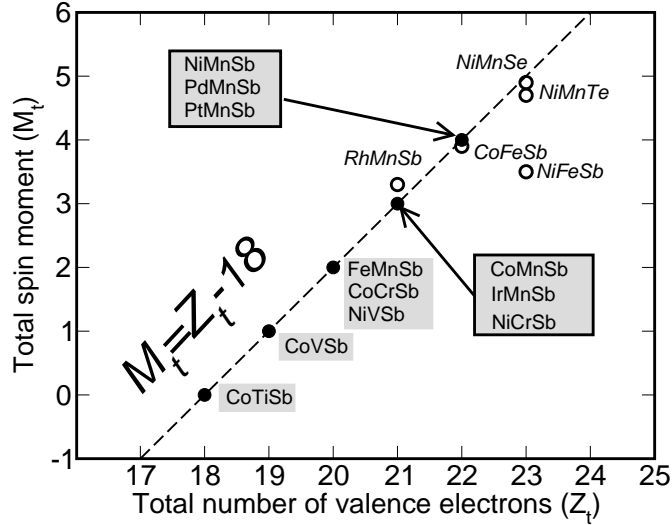}
\caption{Calculated total spin moment per unit cell as a function of
  the total number $Z_t$ of valence electrons per unit cell for all
  the studied half Heusler alloys. The dashed line represents the
  Slater-Pauling behavior. With open circles we present the compounds
  deviating from the SP curve. Some experimental values for bulk
  systems near the SP curve from reference \cite{landolt}: NiMnSb 3.85
  $\mu_B$, PdMnSb 3.95 $\mu_B$, PtMnSb 4.14 $\mu_B$ and finally CoTiSb
  non-magnetic.}
 \label{figios6}
\end{figure}

In figure \ref{figios6} we have gathered the calculated total
spin magnetic moments for the half-Heusler alloys which
we have plotted  as a
function of the total number of valence electrons. The dashed
line represents the rule $ M_t = Z_t -18$ obeyed by these
compounds. The total moment  $M_t$ is an integer
quantity, assuming the values 0, 1, 2, 3, 4 and 5 if $Z_t \ge$18.
The value 0 corresponds to the semiconducting phase and the value
5 to the maximal moment when all 5 majority $d$-states are
filled.  Firstly we varied  the valence of the lower-valent
(\textit{i.e.} magnetic) transition metal atom. Thus we substitute
V, Cr and Fe for Mn in the NiMnSb  and CoMnSb compounds using the
experimental lattice constants of the two Mn compounds. For all
these compounds we find that the total spin moment scales
accurately with the total charge and that they all present the
half-metallicity.

As a next test we have substituted Fe for Mn in CoMnSb and NiMnSb,
but both CoFeSb and NiFeSb lose their half-metallic
character. In the case of NiFeSb the majority $d$-states are
already fully occupied as in NiMnSb, thus the additional electron
has to be screened by the minority $d$-states, so that the Fermi
level falls into the minority Fe states and the half-metallicity
is lost; for half-metallicity a total moment of 5 $\mu_B$ would be
required which is clearly not possible. For CoFeSb the situation
is more delicate. This system has 22 valence electrons and if it
would be a half-metal, it should have a total spin-moment of
4 $\mu_B$ as NiMnSb. In reality our calculations indicate that
the Fermi level is slightly above the gap and the total
spin-moment is slightly smaller than 4 $\mu_B$. The Fe atom
possesses a comparable spin-moment in both NiFeSb and CoFeSb
compounds contrary to the behavior of the V, Cr and Mn atoms.
Except NiFeSb other possible compounds with 23 valence electrons
are NiMnTe and NiMnSe. We have calculated their magnetic
properties using the lattice constant of NiMnSb. As shown in
figure \ref{figios6}, NiMnSe almost makes the 5 $\mu_B$ (its
total spin moment is 4.86 $\mu_B$) and
 is nearly half-metallic, while its isovalent, NiMnTe, has a slightly smaller
spin moment. NiMnSe and NiMnTe  show
big changes in the majority band compared to systems with 22
valence electrons as NiMnSb or NiMnAs, since
antibonding $p$-$d$ states, which are usually above $E_F$, are
shifted below the Fermi level, thus increasing the total moment to
 nearly 5 $\mu_B$.

\section{Full Heusler Alloys\label{secios:3}}

\subsection{Electronic Structure of Co$_2$MnZ with Z = Al, Si, Ga, Ge
  and Sn \label{secios:3-1}}

The second family of Heusler alloys, which we discuss, are the
full-Heusler alloys. We consider in particular compounds containing Co
and Mn, as these are the full-Heusler alloys that have attracted most
of the attention.  They are all strong ferromagnets with high Curie
temperatures (above 600 K) and (except Co$_2$MnAl) they show very
little disorder \cite{landolt}. They adopt the $L2_1$ structure shown
in figure \ref{figios1}. Each Mn or $sp$ atom, sitting in an
octahedral symmetry position, has eight Co atoms as first neighbors,
while each Co has four Mn and four $sp$ atoms as first neighbors and
thus the symmetry of the crystal is reduced to the tetrahedral one.
The Co atoms occupying the two different sublattices are chemically
equivalent as the environment of the one sublattice is the same as the
environment of the second one rotated by 90$^\circ$. Although in the
$L2_1$ structure, the Co atoms are sitting on second neighbor
positions, their interaction is important to explain the magnetic
properties of these compounds as we will show in the next section.

\begin{figure}
\centering
\includegraphics[scale=0.6]{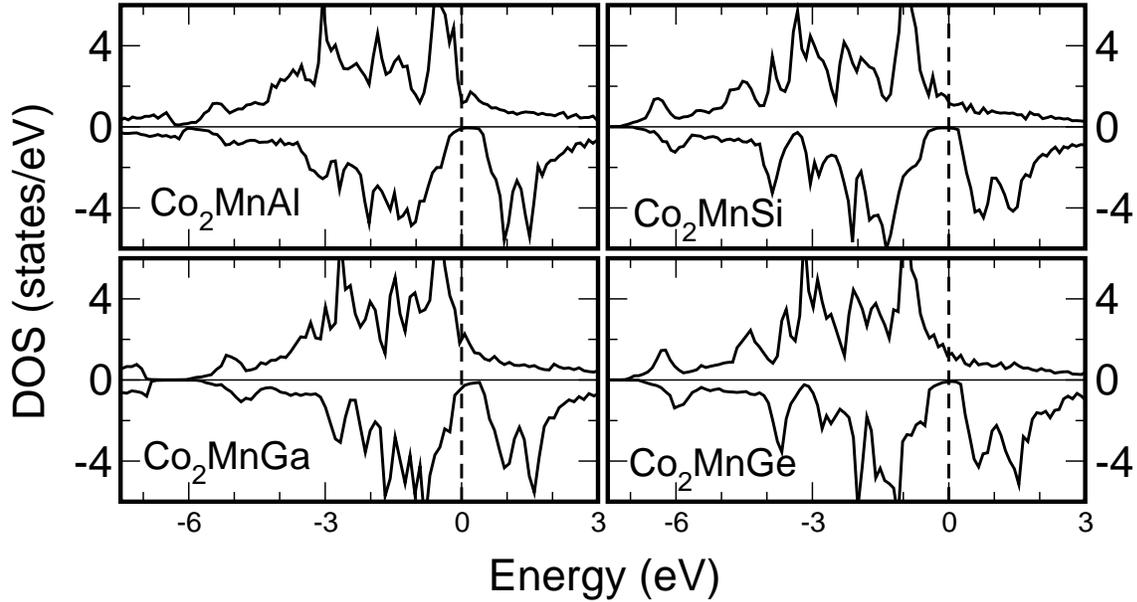}
\caption{Atom-resolved DOS for the Co$_2$MnZ compounds with Z= Al,
Si, Ge, Sn compounds} \label{figios7}
\end{figure}

In figure \ref{figios7} we show the spin-resolved density of states
for the Co$_2$MnAl, Co$_2$MnGa, Co$_2$MnSi and Co$_2$MnGe compounds
calculated using the FSKKR. Firstly as shown by photoemission
experiments by Brown \textit{et al.} in the case of Co$_2$MnSn
\cite{Brown98} and verified by our calculations, the valence band
extends around 6 eV below the Fermi level and the spin-up DOS shows a
large peak just below the Fermi level for these compounds. Ishida
\textit{et al.} \cite{Ishida95} predicted them to be half-metals with
small spin-down gaps ranging from 0.1 to 0.3 eV depending on the
material. Although our previous calculations showed a very small DOS
at the Fermi level, in agreement with the ASW results of K\"ubler et
al. \cite{Kubler84} for Co$_2$MnAl and Co$_2$MnSn, a recalculation of
our KKR results with a higher $\ell$-cut-off of $\ell_{\rm{max}} = 4$
restores the gap and we obtain good agreement with the recent results
of Picozzi et al.~\cite{picozzi} obtained using the FLAPW method.  The
gap is indirect, with the maximum of the valence band at
$\Gamma$ and the minimum of the conduction band at the $X$-point.

In table \ref{tableios1a} we have gathered the atom-resolved and total
moments of the Co$_2$MnZ compounds.  In the case of the half-Heusler
alloys like NiMnSb the Mn spin moment is very localized due to the
exclusion of the spin-down electrons at the Mn site and amounts to
about 3.7 $\mu_B$ in the case of NiMnSb; in CoMnSb the increased
hybridization between the Co and Mn spin-down electrons decreased the
Mn spin moment to about 3.2 $\mu_B$. In the case of the full-Heusler
alloys each Mn atom has eight Co atoms as first neighbors instead of
four as in CoMnSb, and this stronger hybridization is very important
reducing the Mn spin moment even further to less than 3 $\mu_B$,
except in the case of Co$_2$MnSn where it is comparable to the CoMnSb
compound.  The Co atoms are ferromagnetically coupled to the Mn spin
moments and they possess a spin moment that varies from $\sim$0.7 to
1.0 $\mu_B$.  Note that in the half-metallic $C1_b$ Heusler alloys,
the $X$-atom has a very small moment only, in the case of CoMnSb the
Co moment is even negative.  However, in the full Heusler alloys the
Co moment is large and positive and arises basically from two
unoccupied Co bands in the minority conduction band, as explained
below.  Therefore both Co atoms together can have a moment of about 2
$\mu_B$, if all majority Co states are occupied. This is basically the
case for Co$_2$MnSi, Co$_2$MnGe and Co$_2$MnSn (see table
\ref{tableios1a}). In contrast to this, the $sp$ atom has a very small
negative moment which is one order of magnitude smaller than the Co
moment. The negative sign of the induced $sp$ moment characterizes
most of the studied full and half Heusler alloys with very few
exceptions. The compounds containing Al and Ga have 28 valence
electrons and the ones containing Si, Ge and Sn 29 valence electrons.
The former compounds have a total spin moment of 4$\mu_B$ and the
latter ones of 5 $\mu_B$, in agreement with the experimentally deduced
moments of these compounds \cite{Dunlap}. So it seems that the total
spin moment, $M_t$, is related to the total number of valence
electrons, $Z_t$, by the simple relation: $M_t=Z_t-24$, while in the
half-Heusler alloys the total magnetic moment is given by the relation
$M_t=Z_t-18$. In the following section we will analyze the origin of
this rule.
\begin{table}
  \centering \caption{ Calculated spin magnetic moments per unit cell
    in $\mu_B$ for the Co$_2$MnZ compounds, where Z stands for the
    $sp$ atom. (The experimental lattice constants
    \protect{\cite{landolt}} have been used.) The deviations from
    integral total moment for the half-metallic compounds are due to
    numerical angular momentum cutoff (see text). All compounds are
    half-metallic.\label{tableios1a}}
\begin{tabular}{r|r|r|r|r}\hline\noalign{\smallskip}
$m^{spin}$($\mu_B$) &  Co   &    Mn   & Z  & Total \\
\noalign{\smallskip}\hline\noalign{\smallskip}
Co$_2$MnAl    &  0.768  & 2.530 & -0.096 & 3.970  \\
Co$_2$MnGa    &  0.688  & 2.775 & -0.093 & 4.058 \\
Co$_2$MnSi    &  1.021  & 2.971 & -0.074 & 4.940 \\
Co$_2$MnGe    &  0.981  & 3.040 & -0.061 & 4.941 \\
Co$_2$MnSn    &  0.929  & 3.203 & -0.078 & 4.984 \\
\noalign {\smallskip} \hline
\end{tabular}
\end{table}

\subsection{Origin of the gap in Full-Heusler Alloys}
\label{secios:3-2}

Similar to the half-Heusler alloys, the four $sp$-bands are located
far below the Fermi level and are thus irrelevant for the gap.
Therefore, we consider only the hybridization of the 15 $d$ states of
the Mn atom and the two Co atoms. For simplicity we consider only the
$d$-states at the $\Gamma$ point, which show the full structural
symmetry. We will give here a qualitative picture; for a thorough
group theoretical analysis, see reference \cite{Galanakis2002b}. 

\begin{figure}
\centering
\includegraphics[scale=0.7]{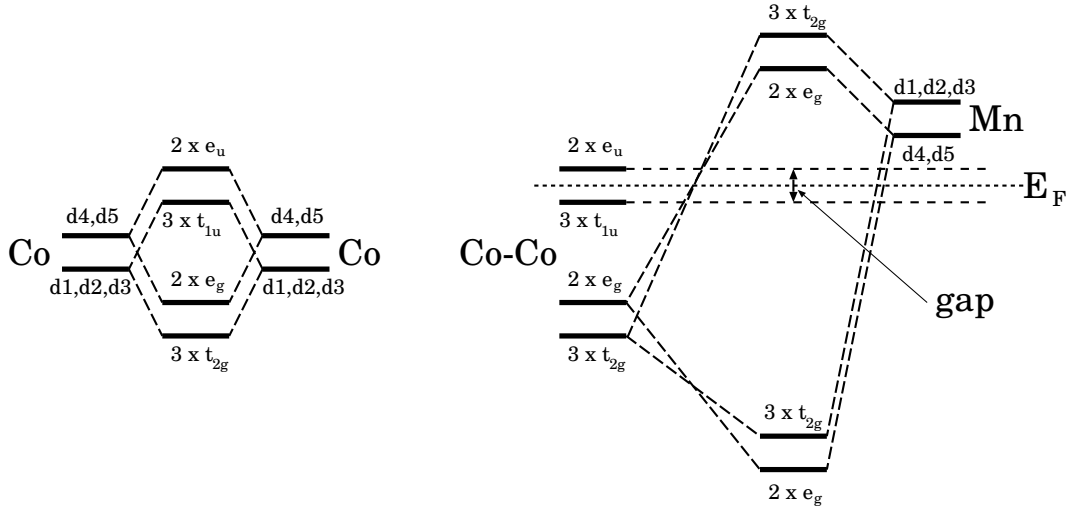}
\caption{Schematic illustration of the origin of the gap in the
minority band in full-Heusler alloys. Due to legibility reasons,
we use d1, d2 and d3 to denote the  $d_{xy}$, $d_{yx}$ and
$d_{zx}$ orbitals, respectively, and d4, d5 for the $d_{r^2}$,
$d_{x^2-y^2}$ orbitals.} \label{figios8}
\end{figure}

Firstly we note that the Co atoms form a simple cubic lattice and that
the Mn atoms (and the Ge atoms) occupy the body centered sites and
have 8 Co atoms as nearest neighbors. Although the distance between
the Co atoms is a second neighbor distance, the hybridization between
these atoms is qualitatively very important.  Therefore we start with
the hybridization between these Co atoms which is qualitatively
sketched in figure \ref{figios8}. The 5 $d$-orbitals are divided into
the twofold degenerate $d_{z^2}$, $d_{x^2-y^2}$ ($e_g$) and the
threefold degenerate $d_{xy}$, $d_{yz}$, $d_{zx}$ ($t_{2g}$) states.
The $e_g$ orbitals ($t_{2g}$ orbitals) of each atom can only couple
with the $e_g$ orbitals $(t_{2g}$ orbitals) of the other Co atom
forming bonding hybrids, denoted by $e_g$ (or $t_{2g}$) and
antibonding orbitals, denoted by $e_u$ (or $t_{1u}$). The coefficients
in front of the orbitals give the degeneracy. Since the Co atoms form
a simple cubic lattice with each Co atom surrounded by 6 other Co
atoms, the crystal field splitting is such that $E_{e_g}>E_{t_{2g}}$.

In a second step we consider the hybridization of these Co-Co orbitals
with the Mn $d$-orbitals. As we show in the right-hand part of figure
\ref{figios8}, the doubly degenerated $e_g$ orbitals hybridize with
the $d_{z^2}$ and $d_{x^2-y^2}$ of the Mn that transform according to
the same representation. They create a doubly degenerate bonding $e_g$
state that is very low in energy and an antibonding one that is
unoccupied and above the Fermi level. The $3\times t_{2g}$ Co orbitals
couple to the $d_{xy,yz,zx}$ of the Mn and create 6 new orbitals, 3 of
which are bonding and occupied and the other three are antibonding and
high in energy. Since each Co atom sits in the center of a Mn
tetrahedron, the crystal field splitting of the bonding and
antibonding states is such that $E_{e_g}<E_{t_{2g}}$. Finally the
$2\times e_u$ and $3\times t_{1u}$ Co orbitals \textit{can not} couple
with any of the Mn $d$-orbitals since none of these is transforming
according to the $u$ representations and since they are orthogonal to
the Co $e_u$ and $t_{1u}$ states. With respect to the Mn and the Ge
atoms these states are therefore non-bonding. The $t_{1u}$ states are
below the Fermi level and they are occupied, while the $e_u$ are just
above the Fermi level. Thus, due to lack of hybridisation with the Mn
states, the octahedral crystal field splitting $E_{e_u}>E_{t_{1u}}$
survives. In summary, in total 8 minority $d$-bands are filled and 7
are empty.

Therefore, all 5 Co-Mn bonding bands are occupied and all 5 Co-Mn
antibonding bands are empty, and the Fermi level falls in-between the
5 non-bonding Co bands, such that the three lower-lying $t_{1u}$ bands
are occupied and the two $e_u$ bands are empty. The maximal moment of
the full Heusler alloys is therefore 7 $\mu_B$ per unit cell, which
would be achieved, if all majority $d$-states were occupied.

\begin{figure}
\centering
\includegraphics[scale=0.6]{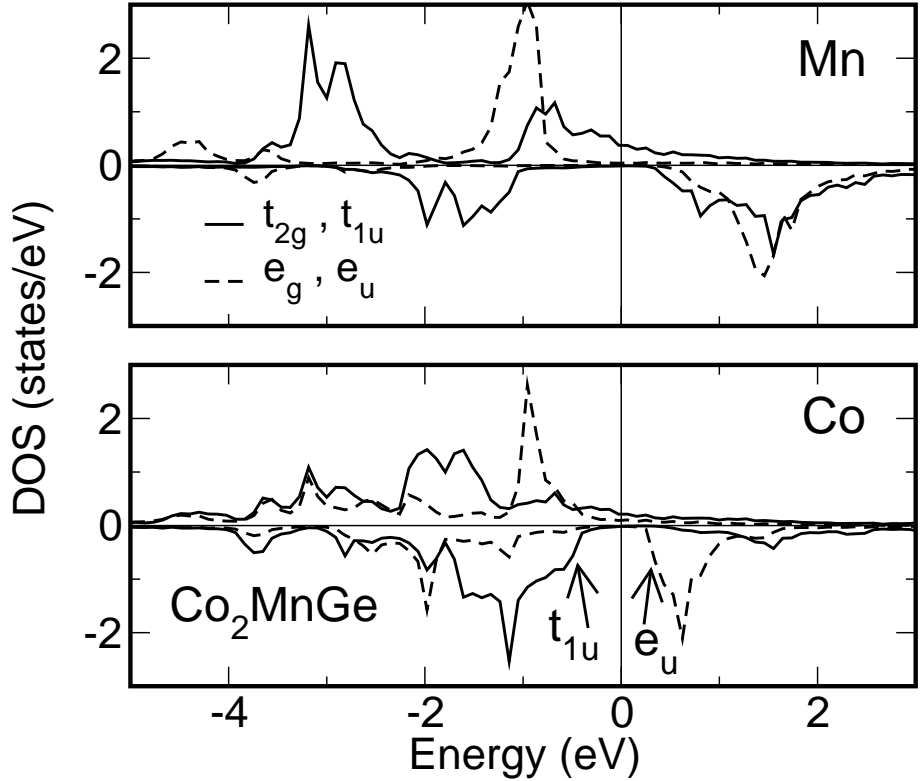}
\caption{Atom- and symmetry-resolved DOS for the
Co$_2$MnGe compound.} \label{figios9}
\end{figure}

In order to demonstrate the existence of the $t_{1u}$ and $e_u$ states
at the Fermi level, we show in figure \ref{figios9} the LDOS of
Co$_2$MnGe at the Co and Mn sites, which are splitted up into the
local $d_{xy}$, $d_{yz}$ and $d_{zx}$ orbitals (normally referred to
as $t_{2g}$; full lines) and the local $d_{z^2}$ and $d_{x^2-y^2}$
orbitals (normally $e_g$; dashed). In the nomenclature used above, the
$d_{xy}$, $d_{yz}$ and $d_{zx}$ contributions contain both the
$t_{2g}$ and the $t_{1u}$ contributions, while the $d_{z^2}$ and
$d_{x^2-y^2}$ orbitals contain the $e_g$ and $e_u$ contributions. The
Mn DOS clearly shows a much bigger effective gap at $E_F$,
considerably larger than in CoMnSb (figure \ref{figios3}), as one
would expect from the stronger hybridization in Co$_2$MnGe. However,
the real gap is determined by the Co-Co interaction only, in fact by
the $t_{1u} - e_u$ splitting, and is smaller than in CoMnSb. Thus the
origin of the gap in the full-Heusler alloys is rather subtle.

\subsection{Slater-Pauling (SP) behavior of the Full-Heusler alloys}
\label{secios:3-3}

\begin{figure}
\centering
\includegraphics[height=7cm]{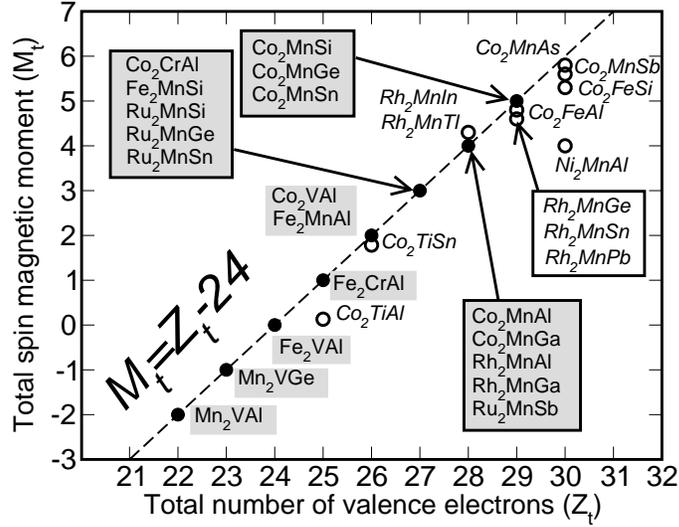}
\caption{Calculated total spin moments for all the studied full
Heusler alloys. The dashed line represents the Slater-Pauling
behavior. With open circles we present the compounds deviating from
the SP curve. Some experimental values for bulk systems near the SP
curve from reference \cite{landolt}: Co$_2$MnAl 4.01 $\mu_B$,
Co$_2$MnSi 5.07 $\mu_B$, Co$_2$MnGa 4.05 $\mu_B$, Co$_2$MnGe 5.11
$\mu_B$, Co$_2$MnSn 5.08 $\mu_B$, Co$_2$FeSi 5.9 $\mu_B$, Mn$_2$VAl
-1.82 $\mu_B$ and finally Fe$_2$VAl non-magnetic.}
 \label{figios10}
\end{figure}

Following the above discussion we investigate the Slater-Pauling
behavior. In figure \ref{figios10} we have plotted the total spin
moments for all the compounds under study as a function of the total
number of valence electrons. The dashed line represents the
half-metallicity rule of the full Heusler alloys: $M_t=Z_t-24$. This
rule arises from the fact that the minority band contains 12 electrons
per unit cell: 4 are occupying the low lying $s$ and $p$ bands of the
$sp$ element and 8 the Co-like minority $d$ bands ($2\times e_g$,
$3\times t_{2g}$ and $3\times t_{1u}$), as explained above (see figure
\ref{figios8}).  Since 7 minority bands (2$\times$Co $e_u$,
5$\times$Mn $d$) are unoccupied, the largest possible moment is 7
$\mu_B$ and occurs when all majority $d$-states are occupied.

Overall we see that many of our results coincide with the
Slater-Pauling curve. Some of the Rh compounds show small deviations
which are more serious for the Co$_2$TiAl compound. We see that there
is no compound with a total spin moment of 7 $\mu_B$ or even 6
$\mu_B$.  Moreover, we show examples of half-metallic materials with
less than 24 electrons, namely Mn$_2$VGe with 23 valence electrons and
Mn$_2$VAl with 22 valence electrons. Firstly, we have calculated the
spin moments of the compounds Co$_2$YAl where Y= Ti, V, Cr, Mn and Fe.
The compounds containing V, Cr and Mn show a similar behavior. As we
substitute Cr for Mn, we depopulate one spin-up state and thus the
spin moment of Cr is around 1 $\mu_B$ smaller than the Mn one while
the Co moments are practically the same for both compounds.
Substituting V for Cr has a larger effect since also the Co spin-up
DOS changes slightly and the Co magnetic moment is increased by about
0.1 $\mu_B$ compared to the other two compounds and V possesses a
small moment of ~0.2 $\mu_B$. This change in the behavior is due to
the smaller hybridization between the Co and V atomic states as
compared to the Cr and Mn ones. Although all three Co$_2$VAl,
Co$_2$CrAl and Co$_2$MnAl compounds are on the SP curve as can be seen
in figure \ref{figios10}, this is not the case for the compounds
containing Fe and Ti. If the substitution of Fe for Mn followed the
same logic as the one of Cr for Mn then the Fe moment should be around
3.5 $\mu_B$ which is a very large moment for the Fe site. Therefore it
is energetically more favorable for the system that also the Co moment
is increased, as was also the case for the other systems with 29
electrons like Co$_2$MnSi, but while the latter one makes it to 5
$\mu_B$, Co$_2$FeAl reaches a value of 4.9 $\mu_B$. In the case of
Co$_2$TiAl, it is energetically more favorable to have a weak
ferromagnet than an integer moment of 1 $\mu_B$ as it is very
difficult to magnetize the Ti atom. Even in the case of the Co$_2$TiSn
the calculated total spin magnetic moment of 1.78 $\mu_B$ (compared to
the experimental value of 1.96 $\mu_B$ \cite{Engen}) arises only from
the Co atoms as was also shown experimentally by Pendl \textit{et al.}
\cite{Pendl}, and the Ti atom is practically nonmagnetic and the
latter compound fails to follow the SP curve.

As a second family of materials we have studied the compounds
containing Fe. Fe$_2$VAl has in total 24 valence electrons and is a
semi-metal, \textit{i.e.} nonmagnetic with a very small DOS at the
Fermi level, as is already known experimentally \cite{Fe2VAl}. All the
studied Fe compounds follow the SP behavior as can be seen in figure
\ref{figios10}. In the case of the Fe$_2$CrAl and Fe$_2$MnAl compounds
the Cr and Mn atoms have spin moments comparable to the Co compounds
and similar DOS. In order to follow the SP curve the Fe in Fe$_2$CrAl
is practically nonmagnetic while in Fe$_2$MnAl it has a small negative
moment. When we substitute Si for Al in Fe$_2$MnAl, the extra electron
exclusively populates Fe spin-up states and the spin moment of each Fe
atom is increased by 0.5 $\mu_B$ contrary to the corresponding Co
compounds where also the Mn spin moment was considerably increased.
Finally we calculated as a test Mn$_2$VAl and Mn$_2$VGe that have 22
and 23 valence electrons, respectively, to see if we can reproduce the
SP behavior for compounds with less than 24 electrons. As we have
already shown, Fe$_2$VAl is nonmagnetic and Co$_2$VAl, which has two
electrons more, has a spin moment of 2 $\mu_B$. Mn$_2$VAl has two
valence electrons less than Fe$_2$VAl and its total spin moment is
$-2\:\mu_B$ and thus it follows the SP behavior; negative total spin
moment means that the ``minority'' band with the gap has more occupied
states than the ``majority'' one.

As we have already mentioned the maximal moment of a full-Heusler
alloy is seven $\mu_B$, and should occur, when all 15 majority $d$
states are occupied. Analogously for a half-Heusler alloy the maximal
moment is 5 $\mu_B$. However this limit is difficult to achieve, since
due to the hybridization of the $d$ states with empty $sp$-states of
the transition metal atoms (sites X and Y in figure \ref{figios1}),
$d$-intensity is transferred into states high above $E_F$, which are
very difficult to occupy. Although in the case of half-Heusler alloys,
we could identify systems with a moment of nearly 5 $\mu_B$, the
hybridization is much stronger in the full-Heusler alloys so that a
total moment of 7 $\mu_B$ seems impossible. Therefore we restrict our
search to possible systems with 6 $\mu_B$, \textit{i.e.} systems with
30 valence electrons, but as shown also in figure \ref{figios10}, none
of them makes the 6 $\mu_B$ exactly. Co$_2$MnAs shows the largest spin
moment: 5.8 $\mu_B$.  The basic reason why moments of 6 $\mu_B$ are so
difficult to achieve, is that as a result of the strong hybridization
with the two Co atoms the Mn atom cannot have a much larger moment
than 3 $\mu_B$. While due to the empty $e_u$-states the two Co atoms
have no problem to contribute a total of 2 $\mu_B$, the Mn moment is
hybridization-limited.

\section{Effect of the Lattice Parameter\label{sect:4}}

\begin{figure}
\centering
\includegraphics[scale=0.6]{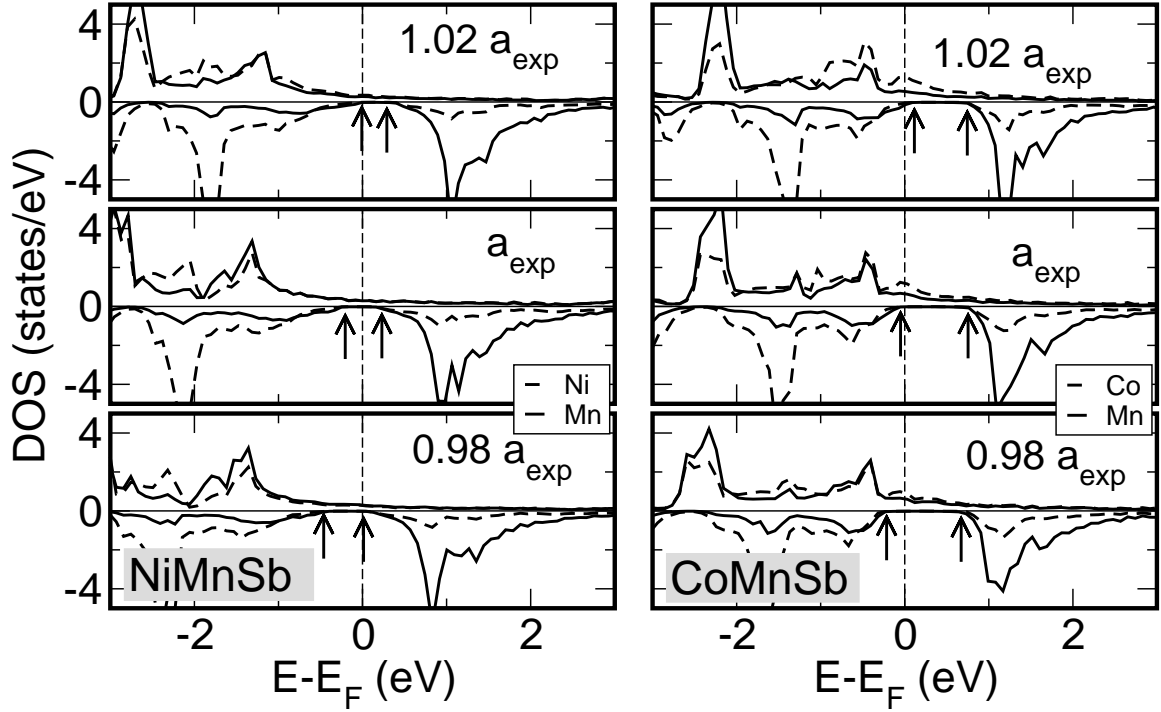}
\caption{Atom-resolved DOS for the experimental lattice parameter
for NiMnSb and CoMnSb, compared with the once compressed or
expanded by 2\%. With the small arrow we denote the edges of the
minority gap.} \label{figios11}
\end{figure}

In this section we will study the influence of the lattice parameter
on the electronic and magnetic properties of the $C1_b$ and $L2_1$
Heusler alloys. To the best of our knowledge no relevant experimental
study exists. For this reason we plot in figure \ref{figios11} the DOS
of NiMnSb and CoMnSb for the experimental lattice parameter and the
ones compressed and expanded by 2 \%. First one sees, that upon
compression the Fermi level moves in the direction of the conduction
band, upon expansion towards the valence band. In both cases, however,
the half-metallic character is conserved. To explain this behavior, we
first note that the Fermi level is determined by the metallic DOS in
the majority band. As we believe, the shift of $E_F$ is determined by
the behavior of the Sb $p$-states, in particular by the large
extension of these states as compared to the $d$ states. Upon
compression the $p$-states are squeezed and hybridize stronger, thus
pushing the $d$-states and the Fermi level to higher energies, i.e.
towards the minority conduction band. In addition the Mn $d$ and Ni or
Co $d$ states hybridize stronger, which increases the size of the gap.
Upon expansion the opposite effects are observed. In the case of
NiMnSb and for the experimental lattice constant the gap-width is
$\sim$ 0.4 eV. When the lattice is expanded by 2\%\ the gap shrinks to
0.25 eV and when compressed by 2\%\ the gap-width is increased to 0.5
eV.  Similarly in the case of CoMnSb, the gap is 0.8 eV for the
experimental lattice constant, 0.65 for the 2\%\ expansion and 0.9 eV
for the case of the 2\%\ compression.

\begin{figure}
\centering
\includegraphics[scale=0.6]{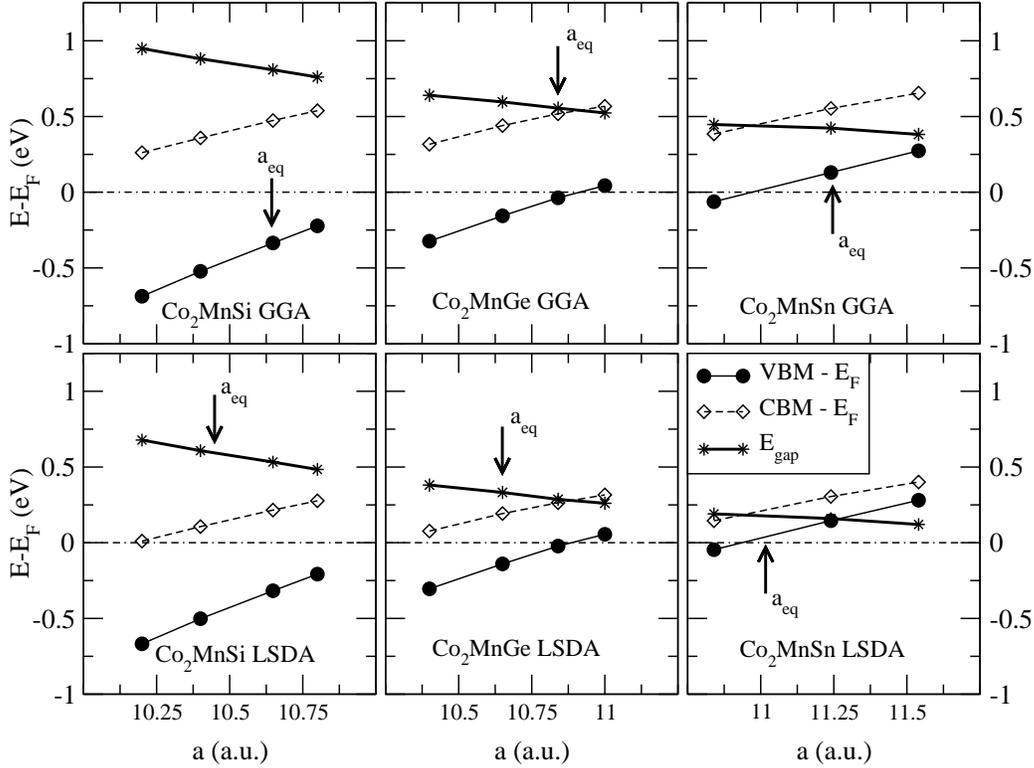}
\caption{Position of the conduction band maximum (CBM) and valence
  band minimum (VBM) relative to the Fermi level for Co$_2$MnSi,
  Co$_2$MnGe, and Co$_2$MnSn, calculated for a range of lattice
  parameters within the LSDA and the GGA. The gap width is also shown.
  The calculated equilibrium lattice parameter is indicated by
  $a_\mathrm{eq}$. (Figure redrawn after Picozzi et al.
  \protect{\cite{picozzi}}.)
  \label{fig:picozzi}}
\end{figure}

For the full-Heusler alloys the pressure dependence has been recently
studied by Picozzi et al. \cite{picozzi} for Co$_2$MnSi, Co$_2$MnGe
and Co$_2$MnSn, using both the LDA and the somewhat more accurate GGA.
The general trends are similar: the minority gap increases with
compression, and the Fermi level moves in the direction of the
conduction band. For example in the case of Co$_2$MnSi the gap-width
is 0.81 eV for the theoretical equilibrium lattice constant of 10.565
\AA . When the lattice constant is compressed to $\sim$ 10.15 \AA, the
gap-width increases to about 1 eV. Results by Picozzi et al.
\cite{picozzi} of are shown in figure~\ref{fig:picozzi}.

The calculations show that for the considered changes of the lattice
constants of $\pm$2 \%, half-metallicity is preserved. There can be
sizeable changes of the local moments, but the total moment remains
constants, since $E_F$ stays in the gap.

\section{Effect of spin-orbit coupling (SOC)\label{sect:5}}

The calculations presented up to this point have neglected the
spin-orbit coupling (SOC). Intuitively, however, one expects that SOC
can be of crucial importance for the half-metallic property: In the
presence of SOC, the electron spin is no more a good quantum number,
so that the electron eigenfunctions cannot conserve their spin degree
of freedom. Wavefunctions within the half-metallic gap must then have
partly spin-down character. As a result, the celebrated half-metallic
gap cannot really be 100\% there even at $T\rightarrow0$. In materials
where the SOC is weak, the DOS within the ``gap'' is expected to be
low, and the polarization close to, but not exacly at, 100\%
\cite{Mavropoulos2004,Mavropoulos2004b}. Another result of SOC is the
appearence of orbital magnetic moment. This, too, is weak in Heusler
alloys with low SOC strength.

\subsection{Polarization in the gap in the presence of SOC}

We begin our discussion with the effect of SOC on the polarization
within the half-metallic gap, where by the term ``gap'' we understand
the energy region where the spin-down DOS is zero if we neglect the
spin-orbit coupling, and very small if we take it into account. The
natural theoretical framework to approach the spin-orbit coupling is
the Dirac equation, where relativistic effects are
inherent. Nevertheless it is more instructive to include SOC as a
perturbation in the Schr\"odinger Hamiltonian, starting from the
unperturbed two-spin-channel description. The perturbation connecting
the two spin channels reads, in terms of the Pauli matrices
$\vec{\sigma}$ and the orbital momentum $\vec{L}$:
\begin{equation}
V_{\mathrm{so}}(r)=\frac{1}{2m^2c^2}\frac{\hbar}{2}
\frac{1}{r}\frac{dV}{dr}\,\vec{L}\cdot\vec{\sigma}
= \left( 
\begin{tabular}{cc}
$V_{\mathrm{so}}^{\uparrow\uparrow}$   & $V_{\mathrm{so}}^{\uparrow\downarrow}$ \\
$V_{\mathrm{so}}^{\downarrow\uparrow}$ & $V_{\mathrm{so}}^{\downarrow\downarrow}$
\end{tabular}
\right).
\label{eq:soc1.0}
\end{equation}
Here, $V(r)$ is the unperturbed one-electron potential at an atomic
site, and is assumed to be spherically symmetric. Deviations from the
spherical symmetry arise only close to the interstisial region between
atoms, where the contribution to the spin-orbit interaction is anyhow
small. The $2\times2$ matrix is the perturbation expressed in spinor
basis, demonstrating the non-diagonal terms
$V_{\mathrm{so}}^{\uparrow\downarrow}$ and
$V_{\mathrm{so}}^{\downarrow\uparrow}$ which are responsible for
spin-flip processes; $\uparrow$ and $\downarrow$ denote the spin up
and spin down direction. Thus, if we denote the unperturbed
hamiltonian for the two spin directions by $H^{0\uparrow}$ and
$H^{0\downarrow}$, and the unperturbed Bloch eigenfunctions as
$\Psi_{n\vec{k}}^{0\uparrow}$ and $\Psi_{n\vec{k}}^{0\downarrow}$,
then the Schr\"odinger equation for the perturbed wavefunction
$\Psi_{n\vec{k}}=(\Psi_{n\vec{k}}^{\uparrow},\Psi_{n\vec{k}}^{\downarrow})$
has the form
\begin{eqnarray}
\left(
\begin{tabular}{cc}
$H^{0\uparrow}+V_{\mathrm{so}}^{\uparrow\uparrow} -E$
& $V_{\mathrm{so}}^{\uparrow\downarrow}$ \\
$V_{\mathrm{so}}^{\downarrow\uparrow}$  &
$H^{0\downarrow}+V_{\mathrm{so}}^{\downarrow\downarrow} -E$
\end{tabular}
\right)
\left(
\begin{tabular}{c}
$\Psi_{n\vec{k}}^{\uparrow}$ \label{eq:2.0}\\
$\Psi_{n\vec{k}}^{\downarrow}$
\end{tabular}
\right)
=0.
\label{eq:soc2.0}
\end{eqnarray}
When $E$ is within the half-metallic band gap, the unperturbed
spin-down solution $\Psi_{n\vec{k}}^{0\downarrow}$ vanishes, and only
the spin-up solution $\Psi_{n\vec{k}}^{0\uparrow}$ is present. Acting
with the perturbed hamiltonian of Eq.~(\ref{eq:soc2.0}) on the
unperturbed spinor $(\Psi_{n\vec{k}}^{0\uparrow},0)$ gives us a
formal solution for the first-order correction to the spin-down
wavefunction:
\begin{equation}
\Psi_{n\vec{k}}^{(1)\downarrow} =
-\frac{1}{H^{0\downarrow}+V_{\mathrm{so}}^{\downarrow\downarrow} - E_{n\vec{k}}^{0\uparrow}} 
V_{\mathrm{so}}^{\downarrow\uparrow} \Psi_{n\vec{k}}^{0\uparrow}
\label{eq:soc3.0}
\end{equation}
where $E_{n\vec{k}}^{0\uparrow}(=E)$ is the energy eigenvalue corresponding
to the state $\Psi_{n\vec{k}}^{0\uparrow}$. This result shows that,
within the gap, the spin-down intensity is a weak image of the band
structure $E_{n\vec{k}}^{0\uparrow}$ of the spin-up band. Since the
spin-down density of states $n_{\downarrow}(E)$ is related to
$|\Psi_{n\vec{k}}^{(1)\downarrow}|^2$, we expect in the gap region a
quadratic dependence of $n_{\downarrow}(E)$ on the spin-orbit coupling
strength: $n_{\downarrow}(E) \sim
(V_{\mathrm{so}}^{\downarrow\uparrow})^2$. Furthermore, the term
$-1/(H^{0\downarrow}+V_{\mathrm{so}}^{\downarrow\downarrow}
-E_{n\vec{k}}^{0\uparrow})$ (this is actually the Green function)
increases the weight of the spin-down states near the $\vec{k}$ points
where the unperturbed spin-up and spin-down bands cross, {\it i.e.},
$E^{0\uparrow}_{n\vec{k}} = E^{0\downarrow}_{n'\vec{k}}$. To see this
we can rewrite Eq.~(\ref{eq:soc3.0}) by expanding the Green function in
spectral representation:
\begin{eqnarray}
\Psi_{n\vec{k}}^{(1)\downarrow}(\vec{r}) &=& 
\int d^3r' \sum_{n'}
\frac{\Psi_{n'\vec{k}}^{(0)\downarrow}(\vec{r})
\Psi_{n'\vec{k}}^{(0)\downarrow *}(\vec{r'})}
{E_{n\vec{k}}^{0\uparrow}-E_{n'\vec{k}}^{0\downarrow}} \,
V_{\mathrm{so}}^{\downarrow\uparrow}(\vec{r'})
\Psi_{n\vec{k}}^{(0)\uparrow}(\vec{r'}) \nonumber\\
&=&
\sum_{n'} 
\frac{\langle\Psi_{n'\vec{k}}^{0\downarrow}|
V_{\mathrm{so}}^{\downarrow\uparrow}|\Psi_{n\vec{k}}^{0\uparrow}\rangle}
{E_{n\vec{k}}^{0\uparrow}-E_{n'\vec{k}}^{0\downarrow}}
\Psi_{n'\vec{k}}^{0\downarrow}(\vec{r})
\label{eq:4.0}
\end{eqnarray}
Here, the summation runs only over the band index $n'$ and not over
the Bloch vectors $\vec{k}'$, because Bloch functions with
$\vec{k}'\neq\vec{k}$ are mutually orthogonal. Close to the crossing
point $E_{n\vec{k}}^{0\uparrow}-E_{n'\vec{k}}^{0\downarrow}$ the
denominator becomes small and the bands strongly couple. Then one
should also consider higher orders in the perturbation
expansion. Such a band crossing can occur at the gap edges, since
within the gap itself there are no spin-down solutions. Therefore, a
blow-up of $n_{\downarrow}(E)$ is expected near the gap edges (but
still within the gap).

The spin polarization $P(E)$ at an energy $E$ (and in particular at
$E_F$) is related to the spin-dependent DOS via the expression
\begin{eqnarray}
P&=&\frac{n_{\uparrow}(E_F)-n_{\downarrow}(E_F)}
{n_{\uparrow}(E_F)+n_{\downarrow}(E_F)} \\
&\approx& 1-2\, n_{\downarrow}(E_F)/n_{\uparrow}(E_F) \ \ \ 
\mbox{for small  $n_{\downarrow}/n_{\uparrow}$ }. \nonumber
\end{eqnarray}

In Refs.~\cite{Mavropoulos2004,Mavropoulos2004b}, $P(E_F)$ was
calculated for a number of half-Heusler alloys and other
half-metals. The calculations were done within density-functional
theory, utilising the KKR Green function method. The approach was
fully relativistic, solving the Dirac equation rather than treating
the spin-orbit coupling as a perturbation. Nevertheless the
qualitative behavior which was described above was
revealed. Table~\ref{table1soc} summarises the results for $P(E_F)$
and for the polarization in the middle of the gap, $P(E_M)$, for the
half-Heusler alloys. The trend is that alloys which include heavier
elements show a lower spin polarization. This becomes most evident by
inspection of the values of $P(E_M)$, since for some of the alloys the
Fermi level approaches or enters the valence band, so that $P(E_F)$
does not always reflect the SOC strength. In particular, NiMnSb shows
a high value for $P(E_M)$ (99.3\%), which decreases when we substitute
the $3d$ element Ni with the heavier, $4d$ element, Pd, and even more
so when we substitute it with the even heavier, $5d$ element, Pt. This
effect is only expected, since it is well-known that heavier elements
show in general a stronger SOC. As an extreme case, the hypothetical
alloy MnBi, which was found to be half-metallic in the zinc-blende
structure if one ignores the spin-orbit interaction~\cite{Xu02}, is
found to have a value of $P(E_F)=77\%$~\cite{Mavropoulos2004b} when
SOC is accounted for (here, the $6p$ wavefunctions of Bi are
resposible for the strength of the effect). As a conclusion, the
spin-orbit coupling reduces $P(E_F)$, but the resulting values are
still high. Nevertheless, we point out that calculations on
half-metallic ferromagnets containing heavy elements ({\it e.g.},
lanthanides) should take into account the spin-orbit coupling for a
reliable quantitative result.
\begin{table}
\begin{center}
\caption{Calculated spin polarisation $P$ at the Fermi level $(E_F)$
  and in the middle of the spin-down gap $(E_M)$, and ratio of
  spin-down/spin-up DOS in the middle of the gap
  $(n_{\downarrow}/n_{\uparrow})(E_M)$, for various half-Heusler
  alloys. The alloys PdMnSb and PtMnSb present a spin-down gap, but
  are not half-metallic, as $E_F$ is slightly below the
  gap. \label{table1soc}}
\begin{tabular}{ccccc}
\hline
Compound & $P(E_F)$ & $P(E_M)$ & $(n_{\downarrow}/n_{\uparrow})(E_M)$ \\
\hline
CoMnSb   &  99.0\%  &   99.5\% & 0.25\% \\
FeMnSb   &  99.3\%  &   99.4\% & 0.30\% \\
NiMnSb   &  99.3\%  &   99.3\% & 0.35\% \\
PdMnSb   &  40.0\%  &   98.5\% & 0.75\% \\
PtMnSb   &  66.5\%  &   94.5\% & 2.70\% \\
\hline
\end{tabular}
\end{center}
\end{table}

\subsection{Orbital moments in Heusler alloys}

The orbital moments in Heusler alloys are expected to be small. This
is a result of the cubic symmetry, of the fact that the magnetic
transition elements here are relatively ligh ($3d$ series) and of the
metallic nature of the electronic states.

Orbital moments for Heusler compounds were calculated by Picozzi {\it
et al.}~\cite{picozzi} for Co$_2$MnSi, -Ge, and -Sn, and more
systematically by Galanakis~\cite{GalanakisOrbit} for ten half-Heusler
and nine full-Heusler compounds. These results are summarized in
tables~\ref{tab:orb1} and~\ref{tab:orb2}. As expected, the values of
the orbital moments $m_{\mathrm{orb}}$ (calculated within the LSDA)
are small. It is known that the LSDA can underestimate
$m_{\mathrm{orb}}$ by up to 50\%, but the trends are considered
reliable. The highest orbital moment, almost 0.1~$\mu_B$, appears at
the Ir and Mn atoms in IrMnSb, but with opposite signs for the two
atoms, so that the total $m_{\mathrm{orb}}$ is close to zero (note
that also the Ir spin moment is opposite to the Mn spin moment; for Ir
in this compound we obtain a ratio of
$m_{\mathrm{orb}}/m_{\mathrm{spin}}\simeq 1/2$).

\begin{table}
\caption{\label{tab:orb1} Spin ($m_\mathrm{spin}$) and orbital
($m_\mathrm{orbit}$) magnetic moments in $\mu_B$ for the XMnSb
half-Heusler compounds. The last three columns are the total spin
and orbital magnetic moment and their sum, respectively}
\begin{tabular}{rrrrrrrrrr}
\hline
\multicolumn{10}{c}{MnSb-based half-Heusler alloys}\\ \hline Alloy&
$m^X_\mathrm{spin}$ & $m^X_\mathrm{orbit}$ &
$m^{Mn}_\mathrm{spin}$ & $m^{Mn}_\mathrm{orbit}$ &
$m^{Sb}_\mathrm{spin}$ & $m^{Sb}_\mathrm{orbit}$ &
$m^{total}_\mathrm{spin}$ & $m^{total}_\mathrm{orbit}$ &
$m^{total}$ \\ \hline FeMnSb  & -0.973&  -0.060&  2.943 &  0.034&
-0.040 &-0.002&  1.958  & -0.028&  1.930\\ CoMnSb  &-0.159&
-0.041& 3.201& 0.032& -0.101&  -0.001  &2.959  & -0.010  &2.949 \\
NiMnSb &0.245& 0.015 &3.720& 0.027  & -0.071&  -0.001 & 3.951  &
0.040 &3.991 \\ CuMnSb & 0.132  & 0.006 &  4.106   &0.032  & 0.028
& -0.006& 4.335  & 0.032   &4.367 \\ RhMnSb &-0.136 &-0.033 &3.627
&0.035& -0.141&  ~ -0& 3.360& 0.001& 3.361 \\ PdMnSb &0.067 &0.007
& 4.036& 0.028 & -0.117& ~ -0& 4.027 &0.035& 4.062 \\ AgMnSb
&0.106 &0.006& 4.334  & 0.031 &  0.040&   -0.007&  4.556 &  0.029
& 4.585 \\ IrMnSb &-0.201&  -0.094& 3.431& 0.092& -0.109 &-0.001
&3.130& -0.004 &3.126 \\ PtMnSb  &0.066& 0.006& 3.911& 0.057&
-0.086 &~ 0& 3.934& 0.063 &  3.997 \\ AuMnSb  &0.134 &  0.021 &
4.335 & 0.027 & 0.056& -0.006&  4.606&   0.044 &  4.650\\
\hline
\end{tabular}
\end{table}

\begin{table}
\caption{\label{tab:orb2}Spin ($m_\mathrm{spin}$) and orbital
($m_\mathrm{orbit}$) magnetic moments in $\mu_B$ for the X$_2$YZ
full-Heusler compounds. The last three colums are the total spin
and orbital magnetic moments and their sum, respectively}
\begin{tabular}{rrrrrrrrrr}
\hline
\multicolumn{10}{c}{Half-metallic full-Heusler alloys}\\
\hline Alloy& $m^X_\mathrm{spin}$ & $m^X_\mathrm{orbit}$ &
$m^Y_\mathrm{spin}$ & $m^Y_\mathrm{orbit}$ & $m^Z_\mathrm{spin}$ &
$m^Z_\mathrm{orbit}$ & $m^{total}_\mathrm{spin}$ &
$m^{total}_\mathrm{orbit}$ & $m^{total}$ \\ \hline Co$_2$MnAl&
0.745 &  0.012 &  2.599 &0.013& -0.091 & ~ -0 & 3.998& 0.038
&4.036
\\ Co$_2$MnSi & 0.994 &0.029& 3.022& 0.017 &  -0.078 & 0.001 &
4.932& 0.076 &  5.008 \\ Co$_2$MnGe&
0.950 &0.030 &3.095 &  0.020 &  -0.065 & 0.001 &  4.931 &0.081  & 5.012 \\
Co$_2$MnSn &0.905& 0.038 &3.257  & 0.025 &  -0.079 & ~  0 &4.988
&0.101 &5.089 \\ Co$_2$CrAl& 0.702& 0.012 & 1.644 &  0.008
&-0.082& ~ 0 &2.966 &  0.033& 2.999
\\  Co$_2$FeAl& 1.094& 0.045  & 2.753 &  0.060 &
-0.095& ~ -0 &4.847 &0.149 &4.996 \\  Fe$_2$MnAl &-0.311 & -0.015
& 2.633& 0.014 &-0.016& 0.001& 1.994 &-0.014 &1.980 \\ Mn$_2$VAl &
-1.398
&-0.034 &0.785& -0.009& 0.013& 0.005& -1.998& -0.073 &-2.071 \\
Rh$_2$MnAl& 0.304& -0.011& 3.431 &0.034 &-0.037 &-0.001 &4.002&
0.011& 4.013\\
\hline
\end{tabular}
\end{table}

\section{Summary and Outlook \label{secios:6}}

We have reviewed the electronic structure and the magnetic properties
of spin-gap Heusler alloys. Although Heusler alloys are complex
systems, calculations based on local density-functional theory
followed by a careful analysis allow reveal the basic mechanisms for
the formation of the gap, as well as the correlation between valence
charge and spin moment, being reflected in the Slater-Pauling
behavior.

The $d$-$d$ hybridization between the transition atoms composing
Heusler alloys is essential for the formation of the gap at $E_F$. In
the case of half-Heusler alloys ({\it e.g.}, NiMnSb) the gap is
created by the hybridization and bonding-antibonding splitting between
the Mn $d$ and the Ni $d$ states. In the case of full-Heusler alloys
({\it e.g.}, Co$_2$MnSi) the gap originates hybridization of the $d$
states of the two Co atoms and the subsequent interaction of these
hybrids with the Mn $d$ states.  At this stage, the $sp$ atom is a
spectator; nevertheless, due to its electronegativity it acts as a
hole reservoir, contributing to a shift of the spin-up bands and
therefore to the value of the total spin moment.  Due to the different
exchange splitting of the different atoms in the unit cell, the
hybridization of spin-up states does not show the same pattern as for
the spin-down states. Thus, in the end, a gap appears only among the
spin-down bands, while for spin-up a metallic character arises.

When the systems are half-metallic, a Slater-Pauling behavior connects
the spin moment $M_t$ per unit cell to the number of valence electrons
$Z_t$ per unit cell by $M_t=(Z_t-18)\ \mu_B$ for half-Heusler alloys
and by $M_t=(Z_t-24)\ \mu_B$ for full-Heusler alloys. These ``magic
numbers'' 18 and 24 (double the number of occupied spin-down valence
states) are consequent of the number and position of the spin-down
states after the $d$-$d$ hybridizations have taken place. The common
nature of the hybridizations among half- (and among full-) Heusler
alloys leads to these Slater-Pauling rules.

While the formation of the gap by hybridization seems straightforward,
the position of $E_F$ with respect to the gap (and thus the
half-metallic property itself) is sensitive to the lattice
constant. Starting from a half-metallic state, compression (or
expansion) of the lattice drives $E_F$ higher into the conduction band
(or lower into the valence band). Thus the half-metallic property,
when present, is stable within 3-5\% change of the lattice parameter. 

The spin-orbit coupling seems to have small effect on the physical
properties of these compounds. On the one hand, the spin polarization
within the gap is reduced, but only by an amount of the order of one
or two percent. On the other hand, the orbital moments are small.

As an outlook we discuss the theoretical perspective of potential
applications of half-metallic Heusler alloys in spintronics. There are
two major issues: one concerning the interfaces of these materials
with semiconductors, and one regarding the half-metallic behavior of
these materials at elevated temperatures (since density-functional
calculations reproduce ground-state properties). On the former
subject, the results so-far are rather discouraging. {\it Ab-initio}
calculations reveal the presence of interface states at almost all
Heusler-semiconductor contacts. These states reduce or even invert the
spin polarization at the interface, an effect which can be critical
for applications~\cite{Mavropoulos2005}. The reason seems to be that,
at the interface with the semiconductor, the $d$-$d$ hybridization
creating the gap stops abruptly. Further research is needed in this
area in order to find material combinations where the hybridization
will continue coherently accross the interface and the gap will be
preserved. Such cases are known to exist for contacts of other types
of half-metals with semiconductors, {\it e.g.}, the half-metallic
zinc-blende pnictides and chalcogenides~\cite{GalanakisZB}.

On the question of elevated temperature, very little theoretical work
has been done~\cite{Dowben03}, due to the complexity that higher
temperature introduces. The high Curie point of Heusler alloys, found
both in experiment and theory~\cite{Tc_heusler}, gives hope for
applicability at room temperature. However, average magnetic order is
not necessarily connected with the presence of a half-metallic gap at
high temperatures. Especially due to their compound nature, Heusler
alloys could have an extremely complex magnetic excitation spectrum
(as is also indicated by experiments~\cite{Hordequin96}), with the
various sublattices exhibiting fluctuations diffrent in nature and
energy scale. Theoretical work in this direction is in progress and
will be reported elsewhere~\cite{Lezaic2005}.

\section*{Acknowledgements}
We would like to thank Silvia Picozzi for useful discussions and for
providing us with her results.


\section*{References}

\begin{thebibliography}{99}


\bibitem{Zutic2004}
\v{Z}uti\'c I,  Fabian J and  Das Sarma A 2004 \RMP \textbf{76}
323

\bibitem{Wolf}
Wolf S A, Awschalom D D,  Buhrman R A,  Daughton J M,  von
Moln\'ar S,  Roukes M L,  Chtchelkanova A Y and  Treger D M 2001
\textit{Science} \textbf{294} 1488

\bibitem{heusler}
Heusler F 1903 \textit{Verh. Dtsch. Phys. Ges.} \textbf{5} 219

\bibitem{landolt}
Webster P J and Ziebeck K R A, in {\em Alloys and Compounds of
d-Elements with Main Group Elements. Part 2.}, edited by H.R.J.
Wijn, Landolt-B\"ornstein, New Series, Group III, Vol. 19,Pt.c
(Springer-Verlag, Berlin), pp. 75-184

\bibitem{landolt2}
Ziebeck K R A and Neumann K -U, in {\em Magnetic Properties of
Metals}, edited by H. R. J. Wijn, Landolt-B\"ornstein, New Series,
Group III, Vol. 32/c (Springer, Berlin), 2001,  pp. 64-414

\bibitem{Pierre97}
Pierre J,  Skolozdra R V,  Tobola J,  Kaprzyk S,  Hordequin C,
Kouacou M A,  Karla I, Currat R and Leli\`evre-Berna E 1997
\textit{J. Alloys Comp.} \textbf{262-263} 101

\bibitem{Tobola}
Tobola J,  Pierre J,  Kaprzyk S,  Skolozdra R V and  Kouacou M A
1998 \JPCM  \textbf{10} 1013; Tobola J and  Pierre J 2000
\textit{J. Alloys Comp.} \textbf{296} 243; Tobola J, Kaprzyk S and
Pecheur P 2003 \PSS (b) \textbf{236} 531

\bibitem{Jussi}
Zayak A T, Entel P, Enkovaara J, Ayuela A, and Nieminen R M 2003
Phys. Rev. B {\bf 68}, 132402

\bibitem{groot}
de Groot R A, Mueller F M, van Engen P G and Buschow K H J 1983
\PRL \textbf{50} 2024

\bibitem{Soulen98}
Soulen Jr R J, Byers J M, Osofsky M S, Nadgorny B, Ambrose T,
Cheng S F, Broussard P R, Tanaka C T, Nowak J, Moodera J S, Barry
A and Coey J M D 1998 \textit{Science} \textbf{282} 85

\bibitem{Kato}
Kato H,  Okuda T,  Okimoto Y,  Tomioka Y,  Oikawa K,
Kamiyama T, and  Tokura T 2004 \PR B {\bf 69}, 184412

\bibitem{Pyrites}
Shishidou T, Freeman A J, and Asahi R 2001 \PR B {\bf 64}, 180401


\bibitem{GalanakisZB}
Galanakis I 2002 \PR B  \textbf{66} 012406; Galanakis I and
Mavropoulos Ph 2003 \PR B  \textbf{67} 104417; Mavropoulos Ph,
Galanakis I and Dederichs P H 2004 \JPCM \textbf{16} 4261

\bibitem{Xie} Sanvito S and Hill N A 2000 \PR B {\bf 62}, 15553;
Continenza A,  Picozzi S,  Geng W T, and  Freeman A J 2001 \PR B
{\bf 64}, 085204; Liu B G 2003 \PR B {\bf 67}, 172411;
Sanyal B,  Bergqvist L, and  Eriksson O 2003 \PR B {\bf 68}, 054417;
 Xie W-H,  Liu B-G, and  Pettifor D G 2003 \PR B {\bf 68},
134407; 
Xie W-H,  Liu B-G, and  Pettifor D G 2003 \PRL
{\bf 91}, 037204;
Xu Y Q,  Liu B G, and
Pettifor D G 2003 \PR B {\bf 68}, 184435; Zhang M et al. 2003
\JPCM {\bf 15}, 5017; 
Fong C Y,  Qian M C,
 Pask J E,  Yang L H, and  Dag S 2004 {\it Appl. Phys. Lett.} {\bf 84}, 239;
Pask J E, Yang L H,  Fong C Y,  Pickett W E, and
Dag S 2003 \PR B {\bf 67}, 224420;
 Zheng J-Cand  Davenport J W 2004 \PR B
{\bf 69}, 144415

\bibitem{Akinaga}
 Akinaga H,  Manago T, and  Shirai M 2000 {\it
  Jpn. J. Appl. Phys.} {\bf 39}, L1118
; 
 Mizuguchi M,  Akinaga H,  Manago T,  Ono K, Oshima M, and
 Shirai M 2002 \JMMM {\bf 239}, 269; 
 Mizuguchi M,
 Akinaga H,  Manago T,  Ono K,  Oshima M,  Shirai M,  Yuri
M,
 Lin H J,  Hsieh H H, and  Chen C T 2002 \JAP {\bf 91}, 7917;
Mizuguchi M,  Ono M K,  Oshima M, Okabayashi J,
 Akinaga H,
 Manago T, and  Shirai M 2002 {\it Surf. Rev. Lett.} {\bf 9}, 331; 
 Nagao,
 Shirai M, and  Miura Y 2004 \JAP {\bf 95}, 6518; 
 Ono,
 Okabayashi J,  Mizuguchi M,  Oshima M,  Fujimori A, and
 Akinaga H 2002 \JAP {\bf 91}, 8088; 
 Shirai M 2001 {\it Physica} E {\bf 10}, 143;
 Shirai M 2003 \JAP {\bf 93}, 6844 

\bibitem{Zhao} Zhao J H, Matsukura F, Takamura K, Abe E, Chiba D, and
  H. Ohno H 2001 {\it Appl. Phys. Lett.} {\bf 79}, 2776; Zhao J H,
  Matsukura F, Takamura K, Abe E, Chiba D, Ohno Y, Ohtani K, and Ohno
  H 2003 {\it Mat. Sci. Semicond. Proc.} {\bf 6}, 507

\bibitem{Temmerman}
 Horne M,  Strange P,  Temmerman W M,  Szotek Z,  Svane A, and
 Winter H 2004 \JPCM {\bf 16}, 5061

\bibitem{FreemanMnSi}
Stroppa A, Picozzi S, Continenza A, and  Freeman A J 2003
\PR B {\bf 68}, 155203

\bibitem{Akai98} Akai H 1998 \PRL {\bf 81}, 3002

\bibitem{Park98}
Park J-H, Vescovo E, Kim H-J, Kwon C,  Ramesh R,
and  Venkatesan T 1998 {\it Nature} {\bf 392}, 794

\bibitem{Datta}
Datta S and Das B 1990 {\it Appl. Phys. Lett.} {\bf 56}, 665

\bibitem{Kilian00}
Kilian K A and Victora R H 2000 \JAP {\bf 87}, 7064

\bibitem{Tanaka99} Tanaka C T,  Nowak J, and Moodera J S 1999
 \JAP {\bf 86}, 6239

\bibitem{Caballero98}  Caballero J A,  Park Y D, Childress
  J R, Bass J,  Chiang W-C,  Reilly A C,  Pratt Jr. W
  P, and Petroff F 1998 {\it J. Vac. Sci. Technol.} A {\bf 16},
  1801 ;  Hordequin C,  Nozi\`eres J P, and Pierre J 1998 \JMMM
  {\bf 183}, 225

\bibitem{Kirillova95}
Kirillova M N, Makhnev A A, Shreder E I, Dyakina V P and Gorina N
B 1995 \PSS (b)  \textbf{187} 231

\bibitem{Hanssen90}
Hanssen K E H M and Mijnarends P E 1990 \PR B \textbf{34} 5009;
Hanssen K E H M, Mijnarends P E, Rabou L P L M and K.H.J. Buschow
K H J 1990 \PR B \textbf{42} 1533

\bibitem{picozziMult}
Picozzi S, Continenza A and  Freeman A J 2003 \JAP \textbf{94}
4723; Picozzi S, Continenza A and  Freeman A J 2003 \textit{J.
Phys. Chem. Solids} \textbf{64} 1697

\bibitem{GalanakisInter2}
Galanakis I 2004 \JPCM \textbf{16} 8007;
Galanakis I, Le\v{z}ai\'c M, Bihlmayer G and Bl\"ugel S 2005 \PR B
{\bf 71}, 214431; Le\v{z}ai\'c M, Galanakis I, Bihlmayer G, and
Bl\"ugel S 2005 \JPCM {\bf 17} 3121


\bibitem{ShiraiInter}
Nagao K, Shirai M and  Miura Y 2004 \JPCM \textbf{16} S5725

\bibitem{Springer}
Galanakis I and Dederichs P H (eds):
Half-metallic alloys: fundamentals and applications, Springer Lecture
notes in Physics vol.~676, Springer (2005).



\bibitem{Zeller95}
Zeller R, Dederichs P H,  \'Ujfalussy B,   Szunyogh L and
Weinberger P 1995  \PR  B \textbf{52} 8807

\bibitem{Papanikolaou02}
Papanikolaou N, Zeller R and Dederichs P H 2002  \JPCM \textbf{14}
2799

\bibitem{Nanda03}
Nanda B R K and Dasgupta I 2003 \JPCM \textbf{15}, 7307


\bibitem{Rudi}
Zeller R 2004 \JPCM  \textbf{16} 6453

\bibitem{Groot86}
de Groot R A,  van der Kraan A M and  Buschow K H J 1986 \JMMM
\textbf{61} 330

\bibitem{Hedin}
Hedin L and Lundqvist S 1969, In: Solid State Physics, vol 23, 
ed by F. Seitz, D. Turnbull, and H. Ehrenreich, (Academic Press, New York and London 1969) pp 1-181

\bibitem{nanda-dasgupta}
Nanda B R K and  Dasgupta S 2003 \JPCM  \textbf{15} 7307

\bibitem{Galanakis2002a}
Galanakis I, Dederichs P H and
  Papanikolaou N 2002 \PR B \textbf{66} 134428

\bibitem{jung}
Jung D,  Koo H J and  Whangbo M J 2000 \textit{J. Mol. Struct.
(Theochem)} \textbf{527} 113


\bibitem{Kubler84}
K\"ubler J 1984 \textit{Physica} B \textbf{127} 257


\bibitem{Brown98}
D Brown, Crapper M D, Bedwell K H, Butterfield M T, Guilfoyle S J,
Malins A E R and Petty M 1998  \PR B \textbf{57} 1563


\bibitem{Ishida95}
Ishida S, Akazawa S, Kubo Y and Ishida J 1982  \JPF \textbf{12}
1111; Ishida S, Fujii S,  Kashiwagi S and Asano S 1995 \JPSJ
\textbf{64} 2152

\bibitem{picozzi}
Picozzi S, Continenza A and  Freeman A J 2002 \PR  B \textbf{66}
094421


\bibitem{Dunlap}
Dunlap R A and Jones D F 1982 \PR B \textbf{26} 6013; Plogmann S,
Schlath\"olter T,  Braun J, Neumann M, Yarmoshenko Yu M,
Yablonskikh M V, Shreder E I, Kurmaev E Z, Wrona A and \'Slebarski
A 1999 \PR  \textbf{60} 6428


\bibitem{Galanakis2002b}
 Galanakis I,  Papanikolaou N and Dederichs P H 2002 \PR B
 \textbf{66} 174429

\bibitem{Engen}
van Engen P G, Buschow K H J  and  Erman M 1983 \JMMM \textbf{30}
374

\bibitem{Pendl}
Pendl Jr W,  Saxena R N,  Carbonari A W,  Mestnik Filho J and
Schaff J 1996 \JPCM  \textbf{8} 11317

\bibitem{Fe2VAl}
Feng Ye, Rhee J Y, Wiener T A,  Lynch D W,  Hubbard
B E,  Sievers A J,  Schlagel D L, Lograsson T A, and
 Miller L L 2001 \PR B {\bf 63}, 165109;  Lue C S,  Ross
Jr. J H ,  Rathnayaka K D D,  Naugle D G,  Wu S Y and
Li W-H 2001 \JPCM {\bf 13}, 1585;  Nishino Y, Kato H,
 Kato M, Mizutani U 2001 \PR B {\bf 63}, 233303; A. Matsushita A,
 Naka T,  Takanao Y,  Takeuchi T, Shishido T, and Yamada Y
2002 \PR B {\bf 65}, 075204

\bibitem{Mavropoulos2004}
Mavropoulos Ph,  Sato K,  Zeller R,  Dederichs P H,  Popescu V and
Ebert H  2004 \PR B \textbf{69} 054424

\bibitem{Mavropoulos2004b} Mavropoulos Ph, Galanakis
I,  Popescu V and Dederichs P H  2004 \JPCM  \textbf{16} S5759

\bibitem{Xu02}
Xu Y-Q, Liu B-G, and  Pettifor D G 2002 \PR B \textbf{66}, 184435


\bibitem{GalanakisOrbit}
Galanakis I 2005 \PR  B \textbf{71} 012413

\bibitem{Mavropoulos2005}
Mavropoulos Ph, Le\v{z}ai\'c M, and Bl\"ugel S 2005 preprint: cond-mat/0506079



\bibitem{Dowben03}
Dowben P A and Skomski R 2003 \JAP \textbf{93}, 7948

\bibitem{Tc_heusler}
Sasioglu E, Sandratskii LM, Bruno P, and Galanakis I 2005 {\it
  Accepted for publication in} \PR B (preprint: cond-mat/0507697)



\bibitem{Hordequin96}
Hordequin Ch, Pierre J, and Currat R 1996 \JMMM {\bf 162}, 75

\bibitem{Lezaic2005}
Le\v{z}ai\'c M, Mavropoulos Ph, Enkovaara J, Bihlmayer G, and Bl\"ugel
S 2005 (unpublished)



\end{thebibliography}
\end{document}